\documentclass[lettersize,journal]{IEEEtran}

\usepackage{amsmath,amsfonts}
\usepackage{booktabs}
\usepackage{algorithmic}
\usepackage{algorithm}
\usepackage{array}
\usepackage[caption=false,font=normalsize,labelfont=sf,textfont=sf]{subfig}
\usepackage{textcomp}
\usepackage{stfloats}
\usepackage{subfig}
\usepackage{bbold}
\usepackage{mathtools}
\usepackage{graphicx}
\usepackage{url}
\usepackage{adjustbox}
\usepackage{verbatim}
\usepackage{graphicx}
\usepackage{multirow}
\usepackage{cite}
\usepackage{hyperref}
\usepackage{cleveref}
\usepackage{xcolor}
\usepackage{adjustbox}
\usepackage{subcaption}

\definecolor{green}{rgb}{0.13, 0.55, 0.13}
\definecolor{red}{rgb}{0.8, 0.0, 0.0}

\begin{document}

\title{Training-free Graph-based Imputation of Missing Modalities in Multimodal Recommendation}

\author{
    \IEEEauthorblockN{Daniele Malitesta\IEEEauthorrefmark{1}, Emanuele Rossi\IEEEauthorrefmark{2}, Claudio Pomo\IEEEauthorrefmark{3}, Tommaso {Di Noia}\IEEEauthorrefmark{3}, Fragkiskos D. Malliaros\IEEEauthorrefmark{1}}\\
    \IEEEauthorblockA{\IEEEauthorrefmark{1}Université Paris-Saclay, CentraleSupélec, Inria, Gif-sur-Yvette, France}\\
    \IEEEauthorblockA{\IEEEauthorrefmark{2}VantAI, Barcelona, Spain}\\
    \IEEEauthorblockA{\IEEEauthorrefmark{3}Politecnico di Bari, Bari, Italy}\\
    \IEEEauthorblockA{Email: \IEEEauthorrefmark{1}name.surname@centralesupelec.fr,
    \IEEEauthorrefmark{2}emanuele.rossi1909@gmail.com,
    \IEEEauthorrefmark{3}name.surname@poliba.it}
}

\markboth{ACCEPTED IN IEEE TRANSACTIONS ON KNOWLEDGE AND DATA ENGINEERING, 2026}%
{Malitesta \MakeLowercase{\textit{et al.}}: Training-free Graph-based Imputation of Missing Modalities in Multimodal Recommendation}

\maketitle

\begin{abstract}
Multimodal recommender systems (RSs) represent items in the catalog through multimodal data (e.g., product images and descriptions) that, in some cases, might be noisy or (even worse) missing. In those scenarios, the common practice is to drop items with missing modalities and train the multimodal RSs on a subsample of the original dataset. To date, the problem of missing modalities in multimodal recommendation has still received limited attention in the literature, lacking a precise formalisation as done with missing information in traditional machine learning. In this work, we first provide a problem formalisation for missing modalities in multimodal recommendation. Second, by leveraging the user-item graph structure, we re-cast the problem of missing multimodal information as a problem of graph features interpolation on the item-item co-purchase graph. On this basis, we propose four training-free approaches that propagate the available multimodal features throughout the item-item graph to impute the missing features. Extensive experiments on popular multimodal recommendation datasets demonstrate that our solutions can be seamlessly plugged into any existing multimodal RS and benchmarking framework while still preserving (or even widen) the performance gap between multimodal and traditional RSs. Moreover, we show that our graph-based techniques can perform better than traditional imputations in machine learning under different missing modalities settings. Finally, we analyse (for the first time in multimodal RSs) how feature homophily calculated on the item-item graph can influence our graph-based imputations.
\end{abstract}


\section{Introduction and motivation}
\label{sec:introduction}

Recommender systems (RSs) are powerful machine learning algorithms designed to derive useful preference patterns of customers over products or services on online platforms (e.g., Amazon, Netflix, Zalando). Despite their large success, mostly built around the collaborative filtering (CF) paradigm, they still fail in inferring users' preferences when the training data is highly sparse~\cite{DBLP:conf/aaai/HeM16}, a phenomenon that occurs quite often in the literature and real-world settings. 

Among popular solutions to address the issue, one of the most effective ways is to leverage products' side information to empower the RSs of useful knowledge regarding users and items in the system. In this respect, multimodal recommender systems~\cite{DBLP:journals/tors/MalitestaCPMNS25, DBLP:journals/csur/LiuHXZGWLT25, DBLP:journals/corr/abs-2302-04473, DBLP:journals/corr/abs-2502-15711, DBLP:conf/kdd/LiuZYDD0ZZD24, DBLP:journals/inffus/PanPCC26} have long settled as the state-of-the-art in personalized recommendation, largely outperforming traditional CF-based recommender systems in providing accurate recommendations. In domains such as fashion~\cite{DBLP:conf/sigir/ChenCXZ0QZ19}, music~\cite{DBLP:conf/recsys/OramasNSS17}, food~\cite{DBLP:journals/tmm/MinJJ20}, and micro-video~\cite{DBLP:journals/tmm/CaiQFX22} recommendation, multimodal RSs leverage multimodal data accompanying items (e.g., product images and descriptions) to extract high-level multimodal features and inject them into the recommendation framework. 

Nevertheless, any multimodal RS seeks \textit{high-quality} multimodal data to derive useful information for the downstream recommendation task, but this requirement cannot always be ensured. In fact, in real-world settings, multimodal data might be noisy or (even worse) \textbf{missing}~\cite{DBLP:conf/cikm/MalitestaRPNM24}. For instance, let us consider a common scenario in e-commerce where, for each product in the catalog, the figure and the description are displayed on the product webpage. It is not rare to encounter situations where the image or also the description are not available. Indeed, the phenomenon is not only limited to real-world settings, as analysing the Amazon Reviews dataset~\cite{DBLP:conf/aaai/HeM16} (\Cref{tab:datasets}) suggests that this also happens in academic research.  

\begin{figure}[!t]
    \centering
    \includegraphics[width=0.9\columnwidth]{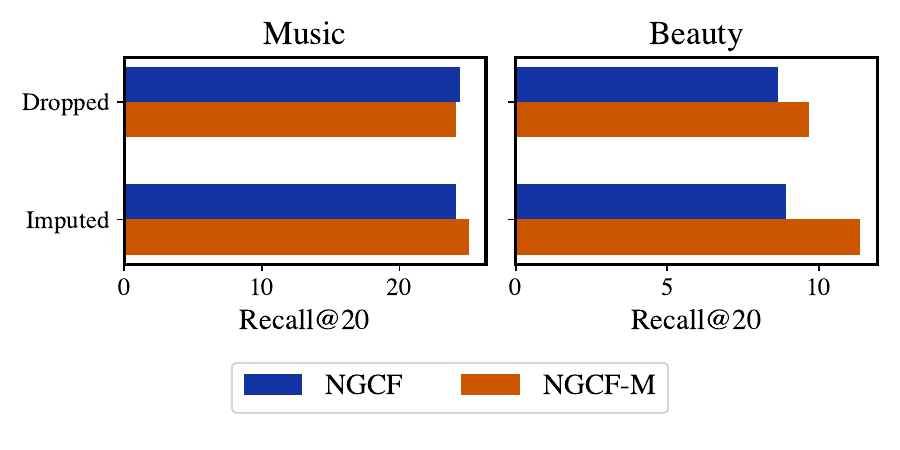}
    \caption{Motivating example where we show the performance improvement (Recall@20) between NGCF (blue) and NGCF-M (brown) in the dropped and imputed settings, for the Music and Beauty datasets. As is evident, the performance improvement of NGCF-M over NGCF becomes wider (or even reverts on the Music dataset) with the imputed setting.}
    \label{fig:motivation}
\end{figure}

In this respect, the standard practice in multimodal recommendation is to drop items with missing modalities and, after this pre-filtering phase, train and evaluate the recommendation model on a subset of the original recommendation data. We maintain that this procedure, despite being common practice in the related literature, becomes \textit{counteractive} to the idea of enhancing the recommendation data through multimodal data to tackle dataset sparsity. To further support the statement, we also show some motivating experimental results (that will be better detailed in \Cref{sec:rq1}). As depicted in~\Cref{fig:motivation}, we report the performance improvement (Recall@20) of a multimodal recommender (NGCF-M~\cite{DBLP:conf/cikm/MalitestaRPNM24}) over its non-multimodal counterpart (NGCF~\cite{DBLP:conf/sigir/Wang0WFC19}) on Music and Beauty categories of the Amazon Reviews dataset, when we drop items with missing modalities and when we impute them (as a pre-processing step). Indeed, we observe that in the \textit{imputed} setting, the performance improvement of NGCF-M over NGCF is much higher (and even reverted) than the one calculated between the same models in the \textit{dropped} setting.
%the recommendation results (Recall@20) on Music and Beauty categories of the Amazon Reviews dataset for two models, NGCF~\cite{DBLP:conf/sigir/Wang0WFC19} and its multimodal version, NGCF-M~\cite{DBLP:conf/cikm/MalitestaRPNM24}, when we drop items with missing modalities and when we impute them (as a pre-processing step). Indeed, we observe that in the \textit{imputed} setting, \textcolor{blue}{the performance of NGCF-M over NGCF is much higher} (and even reverted) than the one calculated between the same models in the \textit{dropped} setting.
This demonstrates that, also empirically, imputing missing modalities in multimodal recommendation can be beneficial to widen the performance gap between traditional and multimodal RSs, as extensively demonstrated in the related literature~\cite{DBLP:conf/aaai/HeM16, DBLP:conf/mm/WeiWN0HC19, DBLP:conf/mm/WeiWN0C20, DBLP:conf/mm/Zhang00WWW21, DBLP:conf/mm/ZhouS23, DBLP:conf/www/ZhouZLZMWYJ23, DBLP:conf/www/WeiHXZ23, DBLP:conf/aaai/GuoL0WSR24, DBLP:conf/aaai/00030000N25}.

On such a basis, it becomes fundamental to design a proper imputation method to recover missing modalities in multimodal recommendation. The problem of missing information in machine learning has been historically discussed and addressed over the past few decades~\cite{DBLP:journals/jbd/EmmanuelMMSMT21}. When it comes to recommendation, several works have tackled the issue of missing information in the user-item interaction data~\cite{DBLP:conf/kdd/Steck10, DBLP:conf/nips/WangGZZ18, DBLP:journals/tkde/ZhengWXLW22} and, in a very few cases, users' and items' metadata~\cite{DBLP:conf/cikm/ShiZYZHLM19, DBLP:journals/tkde/LiuCZLN22}. However, despite the extensive literature on missing modalities in other multimodal learning tasks~\cite{DBLP:conf/aaai/MaRZTWP21, DBLP:conf/cvpr/LeeTCL23, DBLP:conf/kdd/ZhangCMZWWZ22}, the problem has encountered limited (and only recent) attention in multimodal recommendation. The few existing approaches addressing the missingness of modalities in multimodal recommendation~\cite{DBLP:conf/emnlp/WangNL18, DBLP:journals/tmm/YiC22, DBLP:conf/mm/LinTZLW0WY23,  DBLP:conf/recsys/GanhorMHNS24, DBLP:conf/sigir/BaiWHCHZHW24} work within the end-to-end training of the multimodal recommender model by either learning how to impute the missing features or making the recommender system robust to the absence of modalities. However, such solutions are bind to the specific multimodal recommender system they come with, and cannot be applied to any other existing multimodal recommender systems out there. We consider this as a critical limitation since the literature outlines a consistent number of state-of-the-art multimodal recommender systems demonstrating high recommendation performance in standard settings without missing modalities. 

%the recommendation performance (in terms of Recall@20) of NGCF (in blue) and its multimodal version, NGCF-M (in brown), in the dropped and imputed settings, for the Music and Beauty datasets. 

To address the outlined issues, in this work, we propose a family of novel approaches that can be plugged into any multimodal recommender system to impute the missing multimodal features as a pre-processing step before the training of the multimodal recommendation model. Then, unlike the existing solutions, our approaches leverage the graph structure of the user-item recommendation data. Concretely, by considering the item-item co-occurrence graph connecting items that have been interacted by the same users, we propose to re-sketch the problem of missing multimodal features as a graph features imputation problem~\cite{DBLP:conf/log/RossiK0C0B22}. That is, for each modality, we impute the missing features through the available ones by propagating them at multiple hops in the item-item graphs.

Our contributions can be summarized as follows:
\begin{enumerate}
    \item We provide, for the first time in the literature, a problem formalisation for missing modalities in multimodal recommendation, which explicitly outlines differences with missing information in traditional machine learning.
    \item We propose a group of novel imputation approaches that leverage the item-item co-occurrence graph, thus re-interpreting the problem as a task of graph features interpolation. These solutions can be plugged into any existing multimodal RSs and act in a pre-processing manner (i.e., they are \textit{training-free} methods). 
    \item Extensive experiments conducted on six recommendation datasets from the Amazon catalog with 2 modalities, one short-video recommendation dataset with 3 modalities, and 14 (multimodal) recommendation models empirically demonstrate that imputing missing modalities through our approaches can preserve (or even improve) the performance gap between traditional and multimodal RSs. Noticeably, we implement the (multimodal) recommenders either with Elliot~\cite{DBLP:journals/tors/MalitestaCPMNS25}, MMRec~\cite{DBLP:conf/mmasia/Zhou23}, or with their own codebase: this further demonstrates how easy and effective is to apply our imputation pipeline to any already-existing framework in the literature.
    \item Further results prove that our graph-based imputations can outperform traditional imputation methods in machine learning for several settings of missing modalities.
    \item By recognizing the importance that feature homophily~\cite{DBLP:conf/sigir/Zhu0IKF24} (on the item-item graph) can hold on our graph-based imputations, we provide, for the first time in multimodal recommendation with missing modalities, an empirical investigation on their interdependence.
\end{enumerate}

% The remaining sections of the paper are structured as follows. First, in~\Cref{sec:related_work}, we describe the related literature covering the main topics of this work. Second, in~\Cref{sec:background}, we provide the useful background notions that are needed to understand our methodology, later presented in~\Cref{sec:methodology}. After that, in~\Cref{sec:experiments}, we outline the experimental setup for this work, and in~\Cref{sec:results} we report on the results and discuss the main observable trends. Finally, in~\Cref{sec:conclusion}, we summarize the take-home messages and provide interesting directions for the future work. 
Code is accessible at:~\url{https://github.com/danielemalitesta/Graph-Missing-Modalities-TKDE}.
\section{Related work}
\label{sec:related_work}

In this section, we present the related literature on the two main topics for this work: (i) multimodal recommendation and (ii) missing information in recommendation. 
%For each of them, we outline the established and more recent research directions, and describe the limitations that motivated our work.

\subsection{Multimodal recommendation}
In diverse domains such as e-commerce~\cite{DBLP:conf/kdd/Xv0GGLDXZ23},  fashion~\cite{DBLP:conf/sigir/ChenCXZ0QZ19}, music~\cite{DBLP:conf/recsys/OramasNSS17}, food~\cite{DBLP:journals/tmm/MinJJ20}, and micro-video~\cite{DBLP:journals/tmm/CaiQFX22} recommendation, the inclusion of multimodal content associated with items has proven to significantly enhance the representational capability of recommender systems. With the advancements in multimodal learning~\cite{DBLP:books/acm/18/BaltrusaitisAM18}, multimodal recommender systems strive to address persistent challenges in personalized recommendation, such as data sparsity and cold-start~\cite{DBLP:conf/aaai/HeM16,DBLP:conf/recsys/OramasNSS17}. Additionally, exploiting multimodal content can help uncover user-item interactions and intentions through attention mechanisms, thereby enhancing the interpretability of recommendations~\cite{DBLP:conf/sigir/ChenCXZ0QZ19}. In the following paragraphs, we provide a comprehensive overview of the multimodal recommendation domain, as highlighted in recent surveys on the topic~\cite{DBLP:journals/tors/MalitestaCPMNS25, DBLP:journals/csur/LiuHXZGWLT25, DBLP:journals/corr/abs-2302-04473, DBLP:journals/corr/abs-2502-15711, DBLP:conf/kdd/LiuZYDD0ZZD24, DBLP:journals/inffus/PanPCC26}.

Given the recent surge in graph neural networks for recommendation~\cite{DBLP:conf/sigir/0001DWLZ020}, several techniques have begun integrating multimodality into user-item bipartite graphs and knowledge graphs~\cite{DBLP:journals/tmm/WangWYWSN23}. These techniques refine the multimodal representations of users and items through various approaches employing the message-passing schema. While initial efforts involved simply injecting multimodal item features into the graph-based pipeline~\cite{DBLP:conf/mm/WeiWN0HC19}, more advanced methods now learn separate graph representations for each modality and disentangle users' preferences at the modality level~\cite{DBLP:conf/cikm/KimLSK22}, considering both global and local user's interests~\cite{DBLP:conf/aaai/GuoL0WSR24}. Further approaches also concentrate on unveiling multimodal structural distinctions among items in the catalogue~\cite{DBLP:conf/mm/Zhang00WWW21,DBLP:conf/mm/LiuYLWTZSM21, DBLP:conf/mm/ZhouS23, DBLP:conf/ecai/Zhou0Z023}, sometimes leveraging self-supervised~\cite{DBLP:conf/www/ZhouZLZMWYJ23, DBLP:conf/mm/Yu0LB23, DBLP:conf/www/WeiHXZ23, DBLP:conf/aaai/00030000N25, DBLP:journals/tmm/TaoLXWYHC23} and contrastive~\cite{DBLP:conf/sigir/Yi0OM22, DBLP:conf/mm/Su0L0024} learning. Finally, recent works propose to introduce powerful generative approaches such as variational autoencoders~\cite{DBLP:journals/tmm/ZhouM24, DBLP:journals/tmm/YiC24}, diffusion models~\cite{DBLP:conf/mm/JiangX0LLH24}, and large language models~\cite{DBLP:conf/wsdm/WeiRTWSCWYH24}, while others translate the multimodal features to the frequency domain~\cite{DBLP:conf/wsdm/OngK25, DBLP:conf/aaai/Yu0LB25}, transfer the multimodal knowledge from larger to smaller model architectures~\cite{DBLP:conf/apweb/WangYCJYKWHL24, DBLP:conf/www/WeiTXJH24}, or enhance the recommendation of long-tail items from the catalog~\cite{DBLP:conf/cikm/LinMWLZ024}.

In this work, we propose a group of training-free graph-based imputation methods to address missing modalities in multimodal recommendation. Such approaches can be plugged into any multimodal recommender systems presented above. Thus, we select representative (both well-established and recent) multimodal recommender systems and test our solutions on top of them: VBPR~\cite{DBLP:conf/aaai/HeM16}, NGCF-M~\cite{DBLP:conf/cikm/MalitestaRPNM24}, LightGCN-M~\cite{DBLP:conf/www/WeiHXZ23}, FREEDOM~\cite{DBLP:conf/mm/ZhouS23}, BM3~\cite{DBLP:conf/www/ZhouZLZMWYJ23}, MGCN~\cite{DBLP:conf/mm/Yu0LB23}, MMSSL~\cite{DBLP:conf/www/WeiHXZ23}, LGMRec~\cite{DBLP:conf/aaai/GuoL0WSR24}, DiffMM~\cite{DBLP:conf/mm/JiangX0LLH24}, and MENTOR~\cite{DBLP:conf/aaai/00030000N25}.

\subsection{Missing information in recommendation}

In a broader sense, the issue of missing information has always been debated in recommendation. The literature outlines two main research directions regarding missing information on (i) user-item feedback and (ii) content/metadata.

Indubitably, the missing feedback issue is widely recognised as a crucial challenge. On the one hand~\cite{DBLP:journals/siamrev/Strawderman89}, the underneath assumption is that ratings in real-world recommendation data are missing at random since the missing probability does not depend on the ratings themselves; in such a context, learning from observed (not-missing) data is sufficient to provide optimal predictions. On the other hand~\cite{DBLP:journals/siamrev/Strawderman89}, a more complex and interesting scenario is the one of missing not at random ratings, where multiple mechanisms determine the missing feedback ratings; hence, the recommendation algorithm should not ignore these mechanisms during the training. Indeed, the latter setting has been largely investigated over the last few decades~\cite{DBLP:conf/kdd/Steck10,DBLP:conf/nips/WangGZZ18, DBLP:journals/tkde/ZhengWXLW22}, for instance in multicriteria~\cite{DBLP:conf/nss/Takasu11}, social~\cite{DBLP:conf/icdm/Chen0ESFC18}, and movies recommendation~\cite{DBLP:conf/dsaa/VernadeC15}.

Conversely, the issue of missing content in recommendation has been poorly explored in the related literature so far. Based on a careful review, the two principal application scenarios involve items' metadata~\cite{DBLP:conf/cikm/ShiZYZHLM19, DBLP:journals/tkde/LiuCZLN22}. Such works~\cite{DBLP:conf/cikm/ShiZYZHLM19, DBLP:journals/tkde/LiuCZLN22} address the problem of missing attribute entries in the users' and items' metadata following two different strategies, namely, feature sampling towards model's robustness~\cite{DBLP:conf/cikm/ShiZYZHLM19} and graph convolutional networks propagating features on a tripartite graph of users, items, and attributes nodes~\cite{DBLP:journals/tkde/LiuCZLN22}. 

Finally, we consider the problem of missing multimodal information in multimodal recommendation~\cite{DBLP:conf/emnlp/WangNL18, DBLP:journals/tmm/YiC22, DBLP:conf/mm/LinTZLW0WY23,  DBLP:conf/recsys/GanhorMHNS24, DBLP:conf/sigir/BaiWHCHZHW24}, that may fall into the second categorization we made from above. To the best of our knowledge, the literature regarding this aspect is very limited and quite recent~\cite{DBLP:conf/cikm/MalitestaRPNM24}. The solution presented in~\cite{DBLP:conf/emnlp/WangNL18} is one of the first approaches, and works through a generative technique which is trained to reconstruct specific missing modalities embeddings; the only modalities involved are visual and textual. Second, the authors from\cite{DBLP:journals/tmm/YiC22} design a train a novel multimodal model making it robust to the absence of modalities through modality-specific variational autoencoders. Then, the approach proposed in~\cite{DBLP:conf/mm/LinTZLW0WY23} leverages inter- and intra-modality views to impute missing modalities and combine the two outcomes into what pseudo-features replacing the missing ones. More recently, the works in~\cite{DBLP:conf/recsys/GanhorMHNS24, DBLP:conf/sigir/BaiWHCHZHW24} exploit single-branch networks and invariant learning to address missing modalities, but they to not perform any actual multimodal imputation within their pipeline.

Unlike the outlined works, our proposed solutions show several advantages: (i) they act during the pre-processing phase, being completely model-agnostic and avoiding any end-to-end training within the recommendation pipeline; (ii) they leverage information coming from the iter-item co-purchase graph, thus exploiting, for the first time in the related literature, graph-based imputation techniques. Note that, since our approaches are conceptually and technically different from the related works in the literature (i.e., operate during the pre-processing instead of being trained end-to-end), we purposely decide not to compare them one another, as their comparison would be unfeasible and unfair for the above reasons.
\section{Background notions}
\label{sec:background}

In this section, we provide useful background notions for our paper and methodology. Specifically, we will formalise the problem of missing information in machine learning and present popular methods to address it. Then, we will formalise the multimodal recommendation task. 

% In the next paragraphs, we will adopt the following notation. We will use \textbf{boldface} to indicate matrices and tensors, capital letters for matrix/tensor dimensions, and the \textsc{SmallCapital} font for functions. Regarding matrices and tensors, we will use lowercase subscripts to retrieve specific views, where ``*" refers to all values along a specific dimension. Moreover, given a generic matrix/tensor $\mathbf{A}$, we will use the operator $\mathbf{A}[\cdot]$ to retrieve a view of $\mathbf{A}$ through a condition mask (such as a binary mask). 

\subsection{Missing information in machine learning}
\label{sec:missing_machine_learning}

The problem of missing data information in machine learning has been historically discussed for decades~\cite{DBLP:journals/jbd/EmmanuelMMSMT21}.  Without loss of generality, let $\mathbf{X} \in \mathbb{R}^{S \times C}$ be a matrix of $S$ observed samples, each characterised by $C$ features. Then, let $\mathbf{Y} \in \mathbb{R}^{S \times K}$ be a vector mapping each sample to a one-hot encoding class label, where we consider $K$ possible classes. In a multi-class classification problem, we aim to learn a machine learning model (e.g., a neural network) such that it can predict (with a certain approximation error) the class of each label:
\begin{equation}
    \hat{\mathbf{Y}} = \textsc{ML}(\mathbf{X}),
\end{equation}
where $\hat{\mathbf{Y}} \in \mathbb{R}^{S \times K}$ is the predicted vector of labels for each data sample by the learned \textsc{ML}($\cdot$) function.

In real-world settings, the input data may be missing some of (or all) its features. While a possible workaround could be to drop those samples having missing information, another popular procedure involves \textbf{imputing} those missing values in the input data. Let $\mathbf{B} \in \mathbb{R}^{S \times C}$ be a binary matrix such that $\mathbf{B}_{sc} = 1$ if the $c$-th feature value is available for sample $s$, 0 otherwise. Thus, if $\mathbf{B}_{sc} = 0$, we design an \textsc{Impute}($\cdot$) function that approximates the missing value of $c$ for the sample $s$:
\begin{equation}
    \hat{\mathbf{X}}_{sc} = \textsc{Impute}(\mathbf{X}_{*c}[\mathbf{B}_{*c}]),
\end{equation}
where we select, from the input data, all values for feature $c$ and retrieve (with the binary mask and the view operator $[\cdot]$) only the available values and impute the missing one. In general, to impute missing values on the feature $c$, we use the available values from the \textbf{same} feature. We indicate such an imputed value with $\hat{\mathbf{X}}_{sc}$. The \textsc{Impute}($\cdot$) function may be designed in different manners. For instance, we could approximate the missing data with all zero values (i.e., \textsc{Zeros}), with the mean of the available data (i.e., \textsc{GlobalMean}), or random values (i.e., \textsc{Random}).

Once we imputed all missing values from the input data, we train the machine learning model to provide output predictions based on the available and imputed data:
\begin{equation}
     \hat{\mathbf{Y}} = \textsc{ML}(\mathbf{X}[\mathbf{B}], \hat{\mathbf{X}}).
\end{equation}

\subsection{Multimodal recommendation}

Let $\mathcal{U}$ and $\mathcal{I}$ be the sets of users and items, respectively, where $|\mathcal{U}| = U$ and $|\mathcal{I}| = I$. Then, assuming an implicit feedback scenario, we indicate with $\mathbf{R} \in \mathbb{R}^{U \times I}$ the user-item interaction matrix, where $\mathbf{R}_{ui} = 1$ if there exists a recorded interaction between user $u \in \mathcal{U}$ and item $i \in \mathcal{I}$, 0 otherwise.

As in any latent factor-based approaches for recommendation, we map users' and items' IDs to embeddings in the latent space, where $\mathbf{E}_u \in \mathbb{R}^D$ and $\mathbf{E}_i \in \mathbb{R}^D$ represent the $D$-dimensional embeddings for user $u \in \mathcal{U}$ and item $i \in \mathcal{I}$, with $D \ll U, I$. On such a basis, the recommendation task is about reconstructing the user-item interaction matrix $\hat{\mathbf{R}}$ by learning meaningful representations for the user and item embeddings $\mathbf{E}_u$ and $\mathbf{E}_i$ through the existing user-item matrix $\mathbf{R}$: 
\begin{equation}
    \hat{\mathbf{R}} = \textsc{RecSys}(\mathbf{R}, \mathbf{E}_u, \mathbf{E}_i) \quad \forall u \in \mathcal{U}, i \in \mathcal{I}.
\end{equation}

In any multimodal recommendation setting (e.g., fashion, song, micro-video, or food recommendation), items' representation may be suitably enriched by considering the multimodal content describing them, such as products' images, descriptions, or audio tracks. Thus, let $\mathcal{M}$ be the set of modalities (with $|\mathcal{M}| = M$ the number of admissible modalities), and $\mathbf{F} \in \mathbb{R}^{I \times M \times C}$ be the multimodal feature tensor of all items, where $\mathbf{F}_{i*} \in \mathbb{R}^{M \times C}$ is the tensor of all multimodal features of item $i \in \mathcal{I}$, $\mathbf{F}_{*m} \in \mathbb{R}^{I \times C}$ is the tensor of all item features accounting for the modality $m \in \mathcal{M}$, and  $\mathbf{F}_{im} \in \mathbb{R}^{C}$ is the feature of item $i$ accounting for the $m$ modality (e.g., the visual feature extracted from the product image of item $i$ through a pre-trained deep convolutional network~\cite{DBLP:conf/cvpr/HeZRS16}). In both machine learning and multimodal recommendation settings, $C$ refers to data features, standing for data columns (e.g., age, gender, job) in the former, and latent embedding dimension of multimodal features (e.g., 2048 as in the case of ResNet50 to extract visual features from product images) in the latter.  

On such preliminaries, the multimodal recommendation task is about reconstructing the user-item interaction matrix $\hat{\mathbf{R}}$ by learning meaningful user and item embeddings $\mathbf{E}_u$ and $\mathbf{E}_i$ through the existing user-item interaction matrix $\mathbf{R}$ and the multimodal items' features $\mathbf{F}$:
\begin{equation}
\label{eq:mmrecsys}
    \hat{\mathbf{R}} = \textsc{MMRecSys}(\mathbf{R}, \mathbf{E}_u, \mathbf{E}_i, \mathbf{F}) \quad \forall u \in \mathcal{U}, i \in \mathcal{I}.
\end{equation}
\section{Proposed methodology}
\label{sec:methodology}

Unlike other data types in machine learning, recommendation data is modeled by a bipartite and undirected graph representing various relations: user-item, user-user, and item-item. Based on these foundations and  motivated by recent advancements in graph learning~\cite{DBLP:conf/log/RossiK0C0B22}, we propose to frame the problem of imputing missing modalities in recommendation as a problem of \textbf{graph feature interpolation} on the \textbf{item-item} graph. Specifically, we aim to impute missing multimodal features of an item through the existing (i.e., non-missing) multimodal features of the items in the item-item graph. 

\begin{figure*}[!t]
    \centering

    % \subfloat[Traditional imputation]{
    %     \includegraphics[width=0.6\textwidth]{figures/traditional_pipeline.pdf} 
    %     \label{fig:traditiona}
    % }

    % \vspace{1em}
    
    % \subfloat[Graph-based imputation]{
    %     \includegraphics[width=0.85\textwidth]{figures/graph_pipeline.pdf} 
    %     \label{fig:graph}
    % }
    \includegraphics[width=0.9\textwidth]{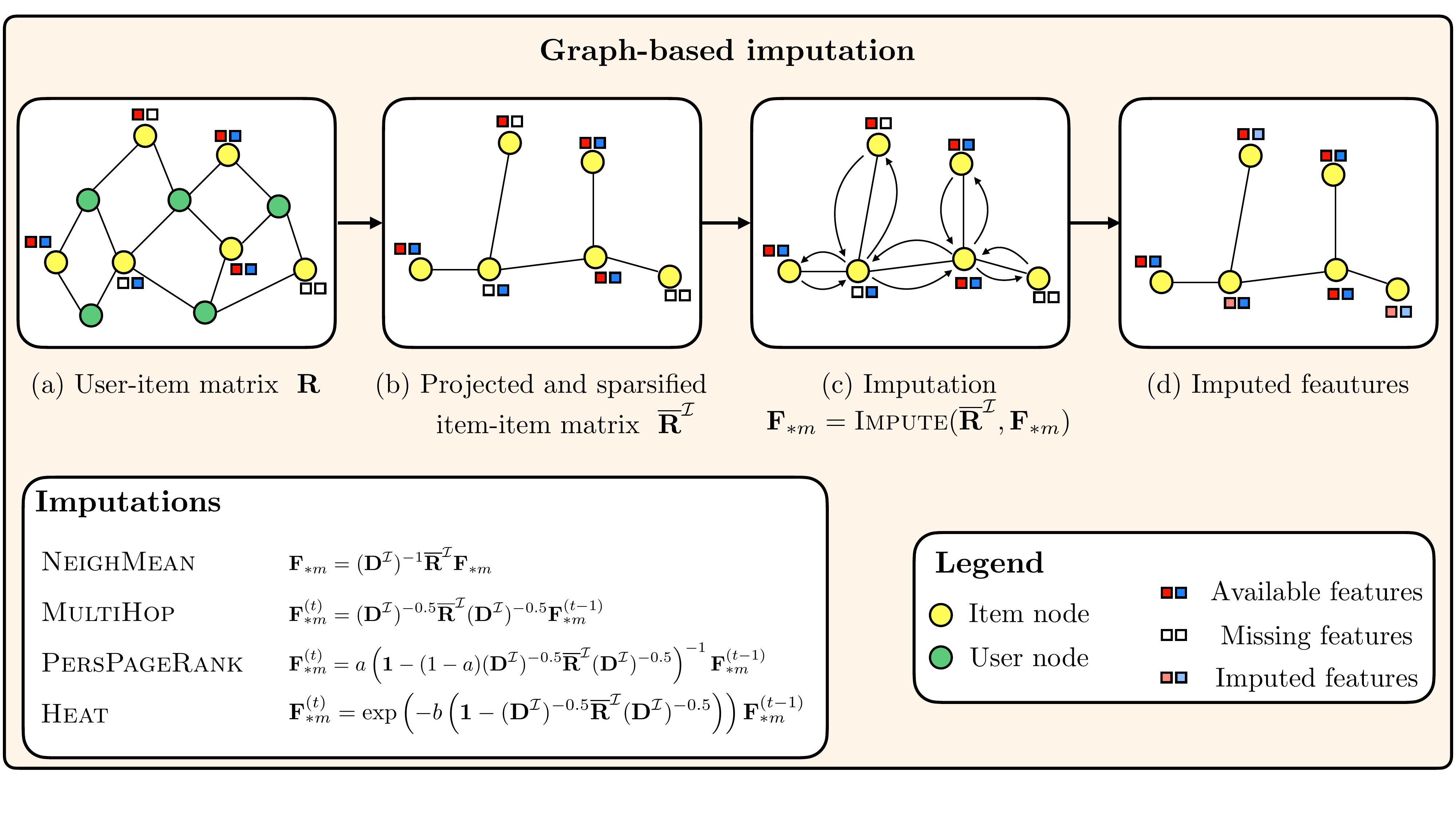}
    % \caption{Visual representation of the differences between imputing with traditional strategies in machine learning and with our graph-based imputations. Above (a), we represent the pipeline for traditional imputation methods, where no relations among items are leveraged. Below (b), we represent the pipeline for our proposed graph-based imputation methods, where we exploit item-item co-occurrences and diffuse the available multimodal features to impute the missing ones.}
    \caption{Visual representation of our graph-based imputations. Starting from the user-item graph with incomplete multimodal features (a), we project and sparsify it to obtain the item-item co-purchase graph (b). Then, we propagate the available multimodal features in the obtained item-item graph (c), which allows us to impute the missing features (d).}
    \label{fig:pipeline}
\end{figure*}

This section is structured as follows. To begin with, it provides a formal definition (for the first time in the literature) of missing modalities in multimodal recommendation. Then, it re-interprets it under the lens of graph feature interpolation. Finally, it presents our pipeline for missing modalities imputation in multimodal recommendation leveraging graph-aware imputation methods. 
%The main differences between our method and traditional imputation in machine learning are highlighted in \Cref{fig:pipeline}.
In \Cref{fig:pipeline}, we visually represent the main stages of our graph-based imputation strategies.

\subsection{Missing modalities in multimodal recommendation}

The problem of missing modalities in multimodal recommendation refers to the unavailability of some (or all) multimodal feature vectors for a specific subset of items (see again~\Cref{tab:datasets}). For instance, when considering a twofold modality setting (e.g., visual and textual), it could happen that item $i$ is missing either the visual or textual modality or both; in every such case, we say item $i$ has some missing modalities. 

The purpose is to design an imputation function \textsc{Impute}($\cdot$) that, similarly to the definition provided in~\Cref{sec:missing_machine_learning}, can approximate the missing multimodal features through the available ones. Nevertheless, unlike traditional machine learning, we consider another version of the binary mask as $\mathbf{B} \in \mathbb{R}^{I \times M \times C}$ such that:
\begin{equation}
    \mathbf{B}_{im} = 
    \begin{cases}
        \mathbf{1}_C & \text{if modality \textit{m} is available for \textit{i}}\\
        \mathbf{0}_C & \text{otherwise},
    \end{cases}
\end{equation}
where $\mathbf{1}_C = [1, 1, \dots, 1] \in \mathbb{R}^C$ and $\mathbf{0}_C = [0, 0, \dots, 0] \in \mathbb{R}^C$. The above outlines that, in the considered setting, the \textbf{whole} feature vector can be available (missing) for item $i$ and modality $m$. Thus, if $\mathbf{B}_{im} = \mathbf{0}_C$, we design an \textsc{Impute}($\cdot$) function that approximates the modality $m$ for item $i$:
\begin{equation}
\label{eq:missing_multimodal}
    \hat{\mathbf{F}}_{im} = \textsc{Impute}(\mathbf{F}_{*m}[\mathbf{B}_{*m}]),
\end{equation}
where we select, from the input multimodal features, those accounting for modality $m$, and retrieve (with a binary mask) only the items with this available modality. Similarly to the definition provided for machine learning, we assume again that we can impute missing multimodal features with the available features from the \textbf{same} modality (e.g., visual with visual, and textual with textual). Nevertheless, in the most general setting, we may want to impute missing multimodal features with \textbf{any} available multimodal features (e.g., visual with textual and vice-versa). We will devote future work to test strategies which apply \textbf{cross-modal} imputation, following similar solutions in the literature which propose the alignment across modalities~\cite{DBLP:conf/aaai/00030000N25, DBLP:conf/www/WeiHXZ23, DBLP:journals/tmm/TaoLXWYHC23, DBLP:conf/www/ZhouZLZMWYJ23}. %While most of the experiments we run involve the former setting, we will dedicate the analysis in~\Cref{sec:rq3} to the latter setting (i.e., \textbf{cross-modal} imputation).

Once we imputed all missing modalities, we train the multimodal recommender system to provide output predictions based on the available and imputed multimodal data:
\begin{equation}
     \hat{\mathbf{R}} = \textsc{MMRecSys}(\mathbf{R}, \mathbf{E}_u, \mathbf{E}_i, \mathbf{F}[\mathbf{B}], \hat{\mathbf{F}}) \; \forall u \in \mathcal{U}, i \in \mathcal{I}.
\end{equation}

\subsection{Graph-aware imputation}
\label{sec:graph-based-proposal}

We propose to frame the problem of imputing missing modalities in multimodal recommendation as a problem of \textbf{graph feature interpolation} on the \textbf{item-item} co-interactions graph. As observed in the literature~\cite{DBLP:conf/log/RossiK0C0B22}, graph feature interpolation requires some prior assumptions on the input graph. One of the common assumptions is the feature \textit{homophily}, which states that the features of each node should be close to those of its neighbors. Translated into our scenario, we assume (with support from the literature~\cite{DBLP:conf/mm/Zhang00WWW21,DBLP:conf/mm/LiuYLWTZSM21, DBLP:conf/mm/ZhouS23}) that \textbf{co-interacted items are likely to present semantically-similar multimodal features}, aligning with the same principle behind collaborative filtering (users sharing similar preferences tend to interact with the same items). In~\Cref{sec:rq4}, we provide a finer-grained evaluation of this aspect, assessing the performance of our approach to different levels of feature homophily~\cite{DBLP:conf/sigir/Zhu0IKF24}.

Starting from the user-item bipartite graph, we build the item-item graph as the projection on the item partition: $\mathbf{R}^{\mathcal{I}} = \mathbf{R}^\top \mathbf{R}$ where $\mathbf{R}^{\mathcal{I}} \in \mathbb{R}^{I \times I}$ and $\mathbf{R}^{\mathcal{I}}_{ij}$ is the number of users who interacted with both items $i$ and $j$, namely, the co-interactions among items from the catalog. As the projected item-item graph may be too dense (i.e., with noisy item-item interactions), we perform the sparsification of the item-item matrix through \textsc{TopN}($\cdot$) sparsification (compare with~\cite{DBLP:conf/mm/Zhang00WWW21}):
\begin{equation}
    \overline{\mathbf{R}}_{ij}^{\mathcal{I}} = \textsc{Sparse}\left(\mathbf{R}^{\mathcal{I}}, N\right) = 
    \begin{cases}
        1 & \text{if } \mathbf{R}_{ij}^{\mathcal{I}} \in \textsc{TopN}\left(\mathbf{R}^{\mathcal{I}}_i\right) \\
        0 & \text{otherwise},
    \end{cases}
\end{equation}
where $\overline{\mathbf{R}}^{\mathcal{I}}$ is the sparsified item-item matrix, while $\textsc{TopN}(\cdot)$ is the function returning the $N$ highest values of a matrix row-wise, with $N$ setting the sparsification rate. Regarding this operation, we acknowledge that the proposed sparsification strategy may lead to possible popularity bias issues where popular items are largely represented in the projected graph; thus, more refined solutions using dynamic thresholding or taking into account interaction frequency might limit the issue~\cite{DBLP:conf/kdd/Xv0GGLDXZ23}. However, our graph-based imputation strategies apply some ad-hoc matrix normalisations, which might effectively prevent this problem (see the following for further details). 

Our goal is to design the graph-aware imputation function \textsc{Impute}($\cdot$) able to approximate the missing multimodal features. Unlike the definition provided in~\Cref{eq:missing_multimodal}, the function takes the available multimodal features \textbf{and} the item-item co-interaction matrix. In other words, the \textsc{Impute}($\cdot$) function approximates the missing multimodal features of an item node through the multimodal features of its neighbors. In general, these latter features may be missing themselves. 

To address this aspect, we design the \textsc{Impute}($\cdot$) function taking as inputs the \textbf{whole} set of multimodal features, and the item-item co-interactions matrix. For the missing multimodal features, and inspired by~\cite{DBLP:conf/log/RossiK0C0B22}, we initialise their embeddings to the zero vector. Thus, to impute the missing features on a given modality $m \in \mathcal{M}$, we do the following:
\begin{equation}
\begin{aligned}
    \text{\footnotesize(i): } & \mathbf{P} = \mathbf{F}_{*m}[\mathbf{B}], \quad \mathbf{F}_{*m}[\neg\mathbf{B}] = \textsc{Zeros}() \\
    \text{\footnotesize(ii): } & \mathbf{F}_{*m} = \textsc{Impute}(\mathbf{\overline{R}}^{\mathcal{I}}, \mathbf{F}_{*m}) \\
    \text{\footnotesize(iii): } & \mathbf{F}_{*m}[\mathbf{B}] = \mathbf{P}, \quad \hat{\mathbf{F}}_{*m} = \mathbf{F}_{*m}[\neg\mathbf{B}],
\end{aligned}
\end{equation}
where we (i) save the available features in a placeholder vector $\mathbf{P}$ and initialise the missing features with the zero vector\footnote{We use the ``$\neg$" operator to indicate the logical negation operator for the binary matrix.}; (ii) impute the missing features with the item-item co-interaction matrix, and \textbf{all} features accounting for that modality; (iii) we extract the imputed missing features and restore the available ones to their original values. In the following, we list four alternatives to design the \textsc{Impute}($\cdot$) function. 

\subsubsection{\textsc{NeighMean}} We extend the \textsc{GlobalMean} imputation through the mean of multimodal features from the one-hop neighborhood (\textsc{NeighMean}):
\begin{equation}
\label{eq:neigh_mean}
    \mathbf{F}_{*m} = 
    %\textsc{NeighMean}(\mathbf{\overline{R}}^{\mathcal{I}}, \mathbf{F}_{*m}) 
    ({\mathbf{D}^{\mathcal{I}}})^{-1} \overline{\mathbf{R}}^{\mathcal{I}} \mathbf{F}_{*m},
\end{equation}
where $\mathbf{D}^{\mathcal{I}} \in \mathbb{R}^{I \times I}$ is the degree matrix for the item-item co-interaction matrix. 

\subsubsection{\textsc{MultiHop}} As we leverage the item-item graph, we can also consider \textbf{multi-hops} relationships between items. Thus, our second graph-aware method iteratively aggregates features from neighborhood nodes (\textsc{MultiHop}):
\begin{equation}
\label{eq:multi-hop}
    \mathbf{F}^{(t)}_{*m} = 
    %\textsc{MultiHop}(\mathbf{\overline{R}}^{\mathcal{I}}, \mathbf{F}^{(t - 1)}_{*m}) \\ 
    ({\mathbf{D}^{\mathcal{I}}})^{-0.5} \overline{\mathbf{R}}^{\mathcal{I}} ({\mathbf{D}^{\mathcal{I}}})^{-0.5} \mathbf{F}^{(t - 1)}_{*m},
\end{equation}
where $\mathbf{F}_{*m}^{(t - 1)}[\mathbf{B}] = \mathbf{P}$ to ensure that, at each new iteration $t \in \{1, \dots, T\}$, we re-initialize the available features to their original values~\cite{DBLP:conf/log/RossiK0C0B22}. Note that we are normalising the aggregated features at each iteration to ensure low-degree nodes are not over-penalised over high-degree ones. This ensures that items initially quite popular in the user-item graph will not be overrepresented during the item-item imputation, thereby limiting the popularity bias problem.

\subsubsection{\textsc{PersPageRank}} While the normalisation operation from above only accounts for the 1-hop neighborhoods, the recent literature has demonstrated that \textbf{graph diffusion at multiple hops}~\cite{DBLP:conf/nips/KlicperaWG19} may improve graph learning; the idea is to leverage an enhanced representation of the interaction matrix, as personalized PageRank~\cite{ilprints422}.

Hence, we propose a third graph-aware imputation that modifies the \textsc{MultiHop} strategy by incorporating \textbf{personalised PageRank} as a normalisation of the item-item co-interactions matrix (\textsc{PersPageRank}):
\begin{equation}
\label{eq:pers_page_rank}
    \mathbf{F}^{(t)}_{*m} = 
    %\textsc{PersPageRank}(\mathbf{\overline{R}}^{\mathcal{I}}, \mathbf{F}^{(t - 1)}_{*m}) \\
    a \left(\mathbb{1} - (1 - a)({\mathbf{D}^{\mathcal{I}}})^{-0.5} \overline{\mathbf{R}}^{\mathcal{I}} ({\mathbf{D}^{\mathcal{I}}})^{-0.5}\right)^{-1}\mathbf{F}^{(t - 1)}_{*m},
\end{equation}
where $a \in [0, 1]$ is the probability of following an edge from the current item or jumping back to a specified set of nodes.

\subsubsection{\textsc{Heat}} With similar rationales to the \textsc{PersPageRank} approach from above, in the \textsc{Heat} strategy we use \textbf{heat diffusion}~\cite{DBLP:conf/icml/KondorL02} for the normalisation of the item-item co-interactions matrix:
\begin{equation}
\label{eq:heat}
    \mathbf{F}_{*m}^{(t)} = 
    %\textsc{Heat}(\mathbf{\overline{R}}^{\mathcal{I}}, \mathbf{F}^{(t - 1)}_{*m}) \\
    \exp\left(-b \left(\mathbb{1} - ({\mathbf{D}^{\mathcal{I}}})^{-0.5} \overline{\mathbf{R}}^{\mathcal{I}} ({\mathbf{D}^{\mathcal{I}}})^{-0.5}\right)\right)\mathbf{F}^{(t - 1)}_{*m},
\end{equation}
where $\exp(\cdot)$ is the exponential function over matrices, and $b$ is the diffusion time. 

\subsection{Differences with item-item recommendation strategies}

Our graph-based imputation solutions present analogies with already-existing multimodal recommendation approaches leveraging item-item similarities, such as~\cite{DBLP:conf/mm/Zhang00WWW21,DBLP:conf/mm/LiuYLWTZSM21, DBLP:conf/mm/ZhouS23, DBLP:conf/ecai/Zhou0Z023}. Thus, it is important to distinguish those methodologies from the proposed ones, evidencing how our graph-based imputations settle as novel techniques in the current literature. We clarify it under two main perspectives.

\begin{itemize}
\item \textbf{The task is different.} Unlike other approaches from the literature, our techniques are specifically tailored to address the imputation of missing modalities in multimodal recommendation. To the best of our knowledge, this is the first attempt in the literature to solve this problem through the graph topology of the user-item dataset.
\item \textbf{The item-item structure is different.} While the resulting item-item similarity graph may appear similar to those proposed in related works, the structure of this graph is technically and conceptually different. The item-item similarity graph implemented in the literature leverages only the multimodal signals, as it is built on the similairities (i.e., cosine similarity) among multimodal embeddings of items. Conversely, in our setting and scenario, some items are missing (part of) their multimodal embeddings, thus it would not be technically and conceptually possible to build an item-item similarity graph through them. That is why we decide to exploit an additional signal, the collaborative one encoded in the user-item dataset, and derive item-item similarities based upon users' interactions (i.e., two items are connected if at least of user has interacted with both of them).
\end{itemize}

\subsection{Further discussion on the proposed imputations}

While effective, our graph-based imputation strategies may, in principle, suffer from inherent limitations stemming from their dependence on the quality of user-item feedback (used to construct the item-item co-interaction graph) and on the availability and reliability of multimodal embeddings, which are exploited to infer missing representations. In this regard, well-known challenges in recommender systems, such as user and item cold-start as well as noisy information, may negatively impact the effectiveness of the proposed imputations. Specifically, items with very limited interactions may be poorly represented in the item-item graph (cold-start), while noisy multimodal embeddings may lead to the unintended propagation of noise during the imputation process.

That said, we emphasize that the primary focus of this work is the problem of missing multimodal information in recommendation, an aspect that remains largely underexplored in both prior and recent literature. We acknowledge that none of the related studies simultaneously address the three challenges discussed here, namely missing multimodal information, cold-start, and noisy data, within a unified framework. Nevertheless, we plan to extend the proposed imputation methodologies to explicitly account for cold-start and noise-related issues, with the goal of ensuring their robustness and effectiveness also in these particularly challenging scenarios.

\begin{table*}[!t]
    \caption{Datasets statistics, where we show items with missing Visual and Textual information, before and after they have been dropped (without and with the apex), and the missing percentage over all the items in the catalog.}\label{tab:datasets}
    \centering
    \begin{tabular}{lrrrrrr}
    \toprule
        \multirow{2}{*}{\textbf{Datasets}} & \multirow{2}{*}{\textbf{$|\mathcal{U}|\;/\;|\mathcal{U}'|$}} & \multirow{2}{*}{\textbf{$|\mathcal{I}|\;/\;|\mathcal{I}'|$}} & \multirow{2}{*}{\textbf{$|\mathcal{R}|\;/\;|\mathcal{R}'|$}} & \multicolumn{3}{c}{\textbf{Missing}} \\ \cmidrule{5-7}
        & & & &  Visual & Textual & Tot. \% \\
        \cmidrule{1-7}
        Office & 4,905 / 4,891 & 2,420 / 1,746 & 53,258 / 35,185 & 0 & 674 & 28\% \\ 
        Music & 5,541 / 5,349 & 3,568 / 2,453 & 64,706 / 51,516 & 2 & 1,114 & 31\% \\
        Baby & 19,445 / 19,440 & 7,050 / 6,382 & 160,792 / 138,149 & 13 & 667 & 9\% \\
        Toys & 19,412 / 19,410 & 11,924 / 11,089 & 167,597 / 146,640 & 17 & 832 & 7\% \\
        Beauty & 22,363 / 22,293 & 12,101 / 11,124 & 198,502 / 165,772 & 7 & 977 & 8\% \\
        Sports & 35,598 / 35,526 & 18,357 / 15,633 & 296,337 / 251,201 & 70 & 2,701 & 15\% \\
        \bottomrule
    \end{tabular}
\end{table*}

\section{Experimental settings}
\label{sec:experiments}

In this section, we describe the experimental settings we adopted in terms of selected recommendation datasets, recommendation baselines and (graph-based) imputation methods, and the reproducibility details to fully replicate our results. 

\subsection{Datasets}

For the sake of this paper, we select six categories of the famous Amazon Reviews dataset~\cite{DBLP:conf/aaai/HeM16}: Office Products (\textbf{Products}), Digital Music (\textbf{Music}), \textbf{Baby}, Toys \& Games (\textbf{Toys}), \textbf{Beauty}, and Sports \& Outdoors (\textbf{Sports}). For each dataset, we consider two versions: (i) \textbf{dropped}, where we remove items with at least one missing modality, along with interactions involving those items and users having interacted only with those items; (ii) \textbf{imputed}, where we impute the items' missing multimodal features. We report the complete statistics of the selected datasets in the two settings, where we distinguish the dropped version from the imputed one through an apex (\Cref{tab:datasets}). From the available items' multimodal data (i.e., product images and textual descriptions), we extract their visual~\cite{DBLP:conf/cvpr/HeZRS16} and textual embeddings~\cite{DBLP:conf/emnlp/ReimersG19}.

\subsection{Baselines}

To test our methodologies, we select recommendation baselines accounting for \textit{traditional} models (without the adoption of multimodal information) and \textit{multimodal} ones. These are: BPRMF~\cite{DBLP:conf/uai/RendleFGS09}, NGCF~\cite{DBLP:conf/sigir/Wang0WFC19}, LightGCN~\cite{DBLP:conf/sigir/0001DWLZ020}, and SGL~\cite{DBLP:conf/sigir/WuWF0CLX21} for the traditional recommendation, and VBPR~\cite{DBLP:conf/aaai/HeM16}, NGCF-M~\cite{DBLP:conf/cikm/MalitestaRPNM24}, LightGCN-M~\cite{DBLP:conf/www/WeiHXZ23}, FREEDOM~\cite{DBLP:conf/mm/ZhouS23}, BM3~\cite{DBLP:conf/www/ZhouZLZMWYJ23}, MGCN~\cite{DBLP:conf/mm/Yu0LB23}, MMSSL~\cite{DBLP:conf/www/WeiHXZ23}, LGMRec~\cite{DBLP:conf/aaai/GuoL0WSR24}, 
DiffMM~\cite{DBLP:conf/mm/JiangX0LLH24}, and MENTOR~\cite{DBLP:conf/aaai/00030000N25} for the multimodal case. Specifically, BPRMF, NGCF, LightGCN, SGL, VBPR, NGCF-M, LightGCN-M, FREEDOM, BM3, and MGCF are implemented in the framework Elliot~\cite{DBLP:journals/tors/MalitestaCPMNS25}, while LGMRec and MENTOR with the framework MMRec~\cite{DBLP:conf/mmasia/Zhou23}, and finally MMSSL and DiffMM with their original codebase. This further demonstrates how our (graph)-based imputations can be seamlessly plugged into any existing multimodal recommendation framework.
As for the imputation approaches, we use \textit{classical machine learning} methods, generative approaches based upon \textit{autoencoders}, and the proposed four \textit{graph-aware} techniques. 

The classical machine learning methods are: (i) \textsc{Zeros} imputes all missing multimodal embeddings to the all-zeros vector. (ii) \textsc{Random} imputes all missing multimodal embeddings to the all-random values vector, by choosing a uniform random distribution. (iii) \textsc{GlobalMean} imputes the missing embeddings from one modality to the average of the available embeddings from the same modality. 

Then, for the autoencoder-based imputations, we select: (i) \textsc{AutoEnc}~\cite{DBLP:conf/aaai/MaRZTWP21} trains an autoencoder to reconstruct missing features from one modality $m_1$ through the available features from another modality $m_2$. The model is trained on the portion of items having both modalities, and then used in inference on the actual missing features.
% Concretely, by retaining the portion of the recommendation data where all items have both modalities, for each item, we train the autoencoder to reconstruct one multimodal embedding through the other (i.e., visual through textual and vice-versa). At inference time, missing visual feautures are reconstructed through their available textual ones and vice-versa. For items where both modalities are missing, we impute the embeddings through the \textsc{Zeros} method. 
(ii) \textsc{GraphAutoEnc} applies a similar approach to \textsc{AutoEnc}, with the difference that it uses a graph-based encoder (i.e., a graph convolutional network~\cite{DBLP:conf/iclr/KipfW17}) operating on the item-item co-purchase graph, obtained through the projection and \textsc{TopN} sparsification from the user-item graph on the item partition. Then, the decoder is again a multi-layer perceptron as in the \textsc{AutoEnc} approach.
% As the model is trained on the portion of recommendation data where all items have both modalities available, we consider the user-item graph that is induced by this set of items and corresponding set of interacting users. 

Finally, our proposed graph-based imputations are (see \Cref{sec:graph-based-proposal}): (i) \textsc{NeighMean}, (ii) \textsc{MultiHop}, (iii) \textsc{PersPageRank}, and (iv) \textsc{Heat}.

\subsection{Reproducibility}

For the training and test of the baselines models, we split the recommendation datasets following the hold-out strategy into train and test sets (80\%/20\%). As for the hyper-parameter search of the recommendation models, we select a shared grid search space with the learning rate in [0.0001, 0.0005, 0.001, 0.005, 0.01] and the regularization term in [1e-5, 1e-2], leaving the other parameters to the best values for each model, and fix the batch size at 1024 and the epochs at 200. Regarding the hyper-parameters for graph-based imputation approaches, we explore the \textsc{TopN} sparsification in [10, 20, \dots, 100] and the propagation hops in [1, 2, \dots, 20]. We use the Recall@20 and the nDCG@20 as test metrics and retain 10\% of the train set as validation, with Recall@20 as a validation metric.

\section{Results and discussion}
\label{sec:results}

In this section, we aim to answer the following research questions (RQs): \textbf{RQ1)} Can the proposed graph-based imputation methods be beneficial in standard multimodal recommendation settings from the literature? \textbf{RQ2)} How do the traditional, autoencoder-, and graph-based imputation methods differently affect the final recommendation performance? \textbf{RQ3)} What is the impact of different sparsification rates and propagation hops on the item-item co-purchase graph for graph-based imputation methods? \textbf{RQ4)} How does the level of feature homophily in the item-item co-purchase graph influence the imputation performance for graph-based imputation approaches? \textbf{RQ5)} What is the performance variation under various percentages of missing modalities for traditional, autoencoders-, and graph-based imputation methods? \textbf{RQ6)} Do the imputation strategies effectively extend to multimodal recommendation datasets with more than two modalities? 

\subsection{RQ1) Graph-based imputations in standard multimodal recommendation}
\label{sec:rq1}
We analyse the impact of our proposed graph-based imputation methods in standard multimodal recommendation settings. We conduct this investigation through two experiments, described in the following. 

In the first investigation, we seek to assess whether and to what extent such imputation methodologies can widen the performance gap between traditional and multimodal recommendation systems. To this end, we select three popular recommendation models, namely, BPRMF, NGCF, and LightGCN, alongside their multimodal versions where we only inject the multimodal items' features (i.e., VBPR, NGCF-M, and LightGCN-M, respectively). We choose these three models because, in their multimodal versions, we do not perform any additional modifications to their original models' architectures apart from the injection of the multimodal features; thus, we can precisely assess the impact of our graph-based imputations of missing modalities and avoid the performance influence from other strategies (e.g., denoising procedures as in FREEDOM). 

\Cref{tab:rq1_1} shows the performance improvement on Recall@20 and nDCG@20 for the three multimodal recommender systems over their original versions, both in the \textbf{dropped} and in \textbf{imputed} settings, where the imputed setting uses the best graph-based imputation according to the Recall@20 calculated on the validation set. We highlight that our analysis focuses on the \textbf{performance improvements} between each non-multimodal recommender and its multimodal counterpart, rather than on the single performance because the dropped and imputed datasets should be regarded, in general, as different datasets (see again \Cref{tab:datasets}), so we could not provide a fair and consistent comparison otherwise. As observable in the table, results largely confirm that the imputed setting not only can preserve the performance gap between traditional and multimodal recommender systems but, in several cases, can widen or even revert it (e.g., consider the case of BPRMF/VBPR on Music or LightGCN/LightGCN-M on Office and Beauty). 

As a confirmation of the outlined outcomes, in our second investigation (i.e., \Cref{tab:rq1_2}), we provide an extensive benchmarking of traditional and multimodal recommender systems from the literature in the dropped setting. Once again, the missing modalities were imputed through the best graph-based imputation method according to the Recall@20 on the validation set. Results generally highlight the performance predominance of multimodal recommender systems over traditional approaches. Moreover, more recent recommendation techniques (e.g., LightGCN and SGL on the traditional side, MMSSL, LGMRec, and MENTOR on the multimodal side) can provide higher recommendation performance.

All things considered, the reported results demonstrate that the application of our proposed graph-based imputation methods for missing modalities can preserve the trends observed in the literature, or even amplify/revert the performance gap between traditional and multimodal approaches. Indeed, the advantages of the imputed setting are to retain important information from the datasets (without the need to perform any filtering/dropping on them) and recover the missing modalities through high-quality strategies. 

\begin{table*}[!t]
\caption{Performance improvement of VBPR, NGCF-M, and LightGCN-M over BPRMF, NGCF, and LightGCN (respectively) in the \textbf{dropped} and \textbf{imputed} settings. For multimodal recommender systems, the selected graph-based imputation method is the one performing the best on the validation set.}\label{tab:rq1_1}
    \centering
    \begin{adjustbox}{width=\textwidth}
    \begin{tabular}{lcccccccccccc}
    \toprule
        \multirow{3}{*}{\textbf{Models}} & \multicolumn{6}{c}{\textbf{Dropped}} & \multicolumn{6}{c}{\textbf{Imputed (Ours)}} \\
        \cmidrule(lr){2-7} \cmidrule(lr){8-13}
        & \multicolumn{2}{c}{\textbf{Office}} & \multicolumn{2}{c}{\textbf{Music}} & \multicolumn{2}{c}{\textbf{Beauty}} &  \multicolumn{2}{c}{\textbf{Office}} & \multicolumn{2}{c}{\textbf{Music}} & \multicolumn{2}{c}{\textbf{Beauty}} \\ \cmidrule(lr){2-3} \cmidrule(lr){4-5} \cmidrule(lr){6-7} \cmidrule(lr){8-9} \cmidrule(lr){10-11} \cmidrule(lr){12-13}
        & Recall & nDCG & Recall & nDCG & Recall & nDCG & Recall & nDCG & Recall & nDCG & Recall & nDCG \\
        \cmidrule{1-13}
        BPRMF & 7.74 & 3.65 & 26.50 & 14.24 & 9.81 & 5.00 & 8.51 & 4.55 & 24.59 & 13.69 & 9.43 & 4.85 \\
        VBPR & 9.99 & 4.82 & 25.22 & 13.71 & 11.12 & 5.86 & 9.47 & 5.23 & 26.25 & 14.58 & 10.92 & 5.71 \\ \cmidrule{1-13}
        Improvement (\%) & \textcolor{green}{+29.07\%} & \textcolor{green}{+32.05\%} & \textcolor{red}{-4.83\%} & \textcolor{red}{-3.72\%} & \textcolor{green}{+13.35\%} & \textcolor{green}{+17.20\%} & \textcolor{green}{+11.28\%} & \textcolor{green}{+14.95\%} & \textcolor{green}{+6.75\%}* & \textcolor{green}{+6.50\%}* & \textcolor{green}{+15.80\%}* & \textcolor{green}{+17.73\%}* \\ \cmidrule{1-13}
        NGCF & 9.90 & 4.49 & 24.39 & 12.61 & 8.64 & 4.28 & 8.18 & 3.91 & 24.12 & 13.06 & 8.92 & 4.56 \\
        NGCF-M & 13.10 & 6.36 & 24.12 & 12.73 & 9.68 & 4.94 & 11.07 & 5.59 & 25.04 & 13.53 & 11.37 & 5.97\\ \cmidrule{1-13}
       Improvement (\%) & \textcolor{green}{+31.32\%} & \textcolor{green}{+41.65\%} & \textcolor{red}{-1.11\%} & \textcolor{green}{+0.95\%} & \textcolor{green}{+12.04\%} & \textcolor{green}{+15.42\%} & \textcolor{green}{+35.33\%}* & \textcolor{green}{+42.97\%}* & \textcolor{green}{+3.81\%}* & \textcolor{green}{+3.60\%}* & \textcolor{green}{+27.47\%}* & \textcolor{green}{+30.92}* \\
       \cmidrule{1-13}
       LightGCN & 13.95 & 6.70 & 27.86 & 14.74 & 11.79 & 6.13 & 11.55 & 6.12 & 26.61 & 14.49 & 11.61 & 6.02 \\
       LightGCN-M & 13.18 & 6.36 & 28.06 & 15.29 & 11.38 & 5.99 & 11.55 & 6.38 & 26.82 & 14.89 & 12.12 & 6.32 \\
       \cmidrule{1-13}
       Improvement (\%) & \textcolor{red}{-5.52\%} & \textcolor{red}{-5.07\%} & \textcolor{green}{+0.72\%} & \textcolor{green}{+3.73\%} & \textcolor{red}{-3.48\%} & \textcolor{red}{-2.28\%} & \textcolor{green}{+0.00\%}* & \textcolor{green}{+4.25\%}* & \textcolor{green}{+0.79\%}* &  \textcolor{green}{+2.76\%} & \textcolor{green}{+4.39\%}* & \textcolor{green}{+4.98\%}* \\
        \bottomrule
        \multicolumn{13}{l}{\footnotesize * \textit{The performance improvement is higher in our imputed setting than in the dropped one.}}
    \end{tabular}
    \end{adjustbox}
\end{table*}

\begin{table*}[!t]
\caption{Recommendation performance in the imputed setting. For multimodal recommender systems, the selected graph-based imputation method is the one performing the best on the validation set. \textbf{Boldface} and \underline{underlined} are the best and second-best values, respectively.}\label{tab:rq1_2}
    \centering
    \begin{adjustbox}{width=0.8\textwidth}
    \begin{tabular}{lcccccccccccc}
    \toprule
        \multirow{2}{*}{\textbf{Models}} &  \multicolumn{2}{c}{\textbf{Office}} & \multicolumn{2}{c}{\textbf{Music}} & \multicolumn{2}{c}{\textbf{Baby}} & \multicolumn{2}{c}{\textbf{Toys}} & \multicolumn{2}{c}{\textbf{Beauty}} & \multicolumn{2}{c}{\textbf{Sports}} \\ \cmidrule(lr){2-3} \cmidrule(lr){4-5} \cmidrule(lr){6-7} \cmidrule(lr){8-9} \cmidrule(lr){10-11} \cmidrule(lr){12-13}
        & Recall & nDCG & Recall & nDCG & Recall & nDCG & Recall & nDCG & Recall & nDCG & Recall & nDCG \\ \cmidrule{1-13}
        BPRMF & 8.51 & 4.55 & 24.59 & 13.69 & 4.56 & 2.23 & 8.55 & 4.55 & 9.43 & 4.85 & 5.32 & 2.73 \\
        NGCF & 8.18 & 3.91 & 24.12 & 13.06 & 5.59 & 2.81 & 8.22 & 4.23 & 8.92 & 4.56 & 4.97 & 2.52 \\
        LightGCN & 11.55 & 6.12 & 26.61 & 14.49 & 7.36 & 3.76 & 10.22 & 5.43 & 11.61 & 6.02 & 8.26 & 4.34 \\
        SGL & 10.28 & 5.81 & 27.59 & 15.86 & 6.08 & 3.27 & 10.89 & 6.05 & 12.20 & 6.72 & 7.87 & 4.27 \\
        \cmidrule{1-13}
        VBPR & 9.47 & 5.23 & 26.25 & 14.58 & 5.90 & 3.03 & 10.36 & 5.63 & 10.92 & 5.71 & 6.65 & 3.39 \\
        NGCF-M & 11.07 & 5.59 & 25.04 & 13.53 & 6.96 & 3.46 & 9.21 & 4.88 & 11.37 & 5.97 & 7.43 & 3.77 \\
        LightGCN-M & 11.55 & 6.38 & 26.82 & 14.89 & 7.72 & 3.91 & 11.24 & 6.10 & 12.12 & 6.32 & 7.50 & 3.89 \\
        FREEDOM & 12.01 & 6.56 & 26.53 & 14.48 & 8.34 & 4.36 & \underline{13.01} & 6.77 & 13.59 & 7.14 & 9.26 & 4.75 \\
        BM3 & 11.20 & 5.94 & 23.99 & 13.10 & 7.89 & 4.01 & 9.87 & 5.17 & 11.51 & 6.00 & 8.55 & 4.44 \\
        MGCN & 11.63 & 6.25 & 24.60 & 13.41 & 8.44 & 4.26 & 12.35 & 6.59 & 12.73 & 6.88 & 8.53 & 4.49 \\
         MMSSL &  \textbf{13.50} &  \textbf{11.38} &  \textbf{29.28} &  \textbf{21.92} &  8.80 &  \textbf{6.55} &  12.04 &  \textbf{8.69} &  13.72 &  \textbf{10.06} &  9.22 &  7.10 \\
         LGMRec &  \underline{13.19} &  \underline{7.28} &  \underline{29.05} &  \underline{17.11} &  \textbf{8.91} &  \underline{4.58} &  12.74 &  6.82 &  \underline{13.76} &  7.56 &  9.23 &  \underline{4.89} \\
         DiffMM &  12.60 &  6.93 &  28.49 &  16.46 &  \underline{8.81} &  4.49 &  11.92 &  6.33 &  13.27 &  7.13 &  \underline{9.29} &  \textbf{4.90} \\
         MENTOR &  12.76 &  7.25 &  27.63 &  15.76 &  8.73 &  4.47 &  \textbf{13.23} &  \underline{7.04} &  \textbf{14.04} &  \underline{7.68} &  \textbf{9.33} &  4.87 \\
        \bottomrule
    \end{tabular}
    \end{adjustbox}
\end{table*}

\subsection{RQ2) Impact of traditional, autoencoder-, and graph-based imputations}

\begin{table*}[!t]
\caption{Impact of all imputation methods on the performance of selected multimodal recommender systems for all the tested datasets. \textbf{Boldface} and \underline{underlined} stand for best and second-best, respectively.}\label{tab:rq2}
    \centering
    \begin{adjustbox}{width=\textwidth}
    \begin{tabular}{llcccccccccccc}
    \toprule
        \multirow{2}{*}{\textbf{Models}} & \multirow{2}{*}{\textbf{Imputation}} & \multicolumn{2}{c}{\textbf{Office}} & \multicolumn{2}{c}{\textbf{Music}} & \multicolumn{2}{c}{\textbf{Baby}} & \multicolumn{2}{c}{\textbf{Toys}} & \multicolumn{2}{c}{\textbf{Beauty}} & \multicolumn{2}{c}{\textbf{Sports}} \\ \cmidrule(lr){3-4} \cmidrule(lr){5-6} \cmidrule(lr){7-8} \cmidrule(lr){9-10} \cmidrule(lr){11-12} \cmidrule(lr){13-14}
        & & Recall & nDCG & Recall & nDCG & Recall & nDCG & Recall & nDCG & Recall & nDCG & Recall & nDCG \\ \midrule
        \multirow{9}{*}{VBPR} & +\textsc{Zeros} & 9.49 & 5.24 & 24.85 & 14.02 & 5.03 & 2.49 & 10.07 & 5.42 & 9.80 & 5.08 & 6.31 & 3.15 \\
        & +\textsc{Random} & 8.57 & 4.75 & 25.65 & 14.24 & 5.24 & 2.64 & 9.80 & 5.37 & 10.09 & 5.22 & 6.52 & 3.30 \\
        & +\textsc{GlobalMean} & 9.46 & 5.30 & 24.81 & 14.03 & 4.98 & 2.46 & 9.76 & 5.33 & 10.55 & 5.55 & 6.31 & 3.15  \\ \cmidrule{2-14}
        & +\textsc{AutoEnc} & 9.34 & 5.02 & 24.76 & 13.95 & 5.01 & 2.48 & 9.45 & 5.09 & 10.13 & 5.33 & 6.30 & 3.13 
        \\ 
        & +\textsc{GraphAutoEnc} & \underline{10.11} & \underline{5.46} & 24.88 & 13.70 & 5.63 & 2.80 & \textbf{10.43} & \textbf{5.68} & 10.87 & \textbf{5.74} & 6.54 & 3.30 \\
        \cmidrule{2-14}
        & +\textsc{NeighMean} & 9.43 & 5.20 & 24.77 & 13.69 & 4.99 & 2.48 & 10.10 & 5.45 & 9.80 & 5.08 & \textbf{6.71} & \textbf{3.42} \\
        & +\textsc{MultiHop} & 10.09 & 5.41 & \underline{26.03} & \underline{14.56} & 5.68 & 2.81 & 10.31 & 5.61 & 10.83 & \underline{5.73} & 6.56 & 3.38 \\
        & +\textsc{PersPageRank} & 9.47 & 5.23 & \textbf{26.25} & \textbf{14.58} & \underline{5.83} & \underline{2.95} & \underline{10.36} & \underline{5.63} & \underline{10.91} & \textbf{5.74} & \underline{6.66} & \underline{3.40} \\
        & +\textsc{Heat} & \textbf{10.21} & \textbf{5.64} & 25.23 & 14.39 & \textbf{5.90} & \textbf{3.03} & 10.35 & \underline{5.63} & \textbf{10.92} & 5.71 & 6.65 & 3.39 \\
        \midrule
        \multirow{9}{*}{FREEDOM} & +\textsc{Zeros} & 11.10 & 5.87 & 25.41 & 13.57 & 7.30 & 3.73 & 9.66 & 5.18 & 10.65 & 5.54 & 7.48 & 3.94 \\
        & +\textsc{Random} & 12.30 & 6.53 & 26.49 & 14.72 & 8.37 & 4.30 & 12.71 & 6.63 & 13.12 & 6.87 & 9.00 & 4.66 \\
        & +\textsc{GlobalMean} & \underline{12.36} & \underline{6.63} & \textbf{26.84} & \textbf{14.92} & 8.25 & 4.23 & 12.52 & 6.53 & 13.08 & 6.80 & 8.89 & 4.63 \\ \cmidrule{2-14} 
        & +\textsc{AutoEnc} & \textbf{12.42} & 6.51 & 25.99 & 14.20 & 7.51 & 3.81 & 9.46 & 5.08 & 12.00 & 6.32 & 7.61 & 4.02 \\ 
        & +\textsc{GraphAutoEnc} & 12.23 & 6.42 & \underline{26.73} & 14.50 & 7.60 & 3.87 & 9.96 & 5.34 & 12.11 & 6.45 & 7.65 & 4.01 \\ \cmidrule{2-14}
        & +\textsc{NeighMean} & 12.30 & 6.45 & 26.57 & 14.55 & \underline{8.63} & \underline{4.42} & 12.69 & 6.67 & 13.23 & 6.88 & 8.96 & 4.63 \\
        & +\textsc{MultiHop} & 12.31 & \textbf{6.68} & 26.53 & 14.48 & 8.54 & 4.39 & 12.86 & \underline{6.78} & \textbf{13.59} & \textbf{7.14} & \underline{9.11} & \underline{4.71} \\
        & +\textsc{PersPageRank} & 12.01 & 6.56 & 26.69 & \underline{14.72} & 8.34 & 4.36 & \underline{12.94} & \textbf{6.79} & 13.28 & 6.98 & \textbf{9.26} & \textbf{4.75} \\
        & +\textsc{Heat} & 12.25 & 6.48 & 26.52 & 14.43 & \textbf{8.68} & \textbf{4.44} & \textbf{13.01} & 6.77 & \underline{13.40} & \underline{7.00} & 8.96 & 4.67 \\
        \midrule
        \multirow{9}{*}{BM3} & +\textsc{Zeros} & 11.31 & 5.93 & 24.64 & 12.99 & 7.64 & 3.87 & 9.75 & 5.02 & 11.31 & 5.92 & 8.43 & 4.43 \\
        & +\textsc{Random} & 10.76 & 5.80 & 24.29 & 12.93 & 7.53 & 3.84 & 9.64 & 5.07 & 11.43 & 5.94 & 8.49 & \underline{4.45}  \\
        & +\textsc{GlobalMean} & 10.93 & 5.79 & 24.27 & 12.88 & 7.70 & 3.94 & 9.84 & 5.13 & 11.37 & 5.87 & \underline{8.56} & \underline{4.45}  \\ \cmidrule{2-14} & +\textsc{AutoEnc} & 10.92 & 5.87 & \textbf{25.04} & \textbf{13.48} & 7.86 & 3.99 & 9.98 & 5.24 & 11.30 & 5.92 & 8.25 & 4.40  \\ 
        & +\textsc{GraphAutoEnc} & 10.86 & 5.92 & 24.53 & 13.00 & 7.85 & \underline{4.00} & \textbf{10.17} & \underline{5.26} & 11.41 & 5.95 & 8.52 & 4.45 \\
        \cmidrule{2-14}
        & +\textsc{NeighMean} & 10.91 & 5.71 & 24.46 & 12.95 & 7.62 & 3.91 & 9.77 & 5.09 & 11.43 & 5.94 & 8.36 & 4.35 \\
        & +\textsc{MultiHop} & \textbf{11.40} & 5.60 & \underline{24.67} & \underline{13.14} & \underline{7.87} & \textbf{4.01} & 9.87 & 5.17 & \underline{11.49} & \textbf{6.04} & 8.46 & 4.42 \\
        & +\textsc{PersPageRank} & 11.20 & \underline{5.94} & 23.99 & 13.10 & \textbf{7.89} & \textbf{4.01} & 9.79 & 5.12 & 11.31 & 5.92 & 8.55 & 4.44 \\
        & +\textsc{Heat} & \underline{11.35} & \textbf{6.04} & 24.41 & 12.97 & \textbf{7.89} & \textbf{4.01} & \underline{10.12} & \textbf{5.27} & \textbf{11.51} & \underline{6.00} & \textbf{8.63} & \textbf{4.49} \\
        \midrule
        \multirow{9}{*}{MGCN} & +\textsc{Zeros} & 11.17 & 5.80 & 20.90 & 10.97 & 5.64 & 2.79 & 5.51 & 2.88 & 10.77 & 5.92 & 5.13 & 2.75 \\
        & +\textsc{Random} & 11.64 & 6.06 & 21.16 & 11.75 & 8.28 & 4.24 & 11.93 & 6.35 & 12.69 & 6.83 & 8.31 & 4.34 \\
        & +\textsc{GlobalMean} & 11.45 & 6.12 & 22.00 & 11.89 & 8.20 & 4.20 & 11.91 & 6.30 & 12.62 & 6.72 & 8.29 & 4.36  \\
        \cmidrule{2-14}
        & +\textsc{AutoEnc} & 11.76 & 6.21 & 23.27 & 12.91 & 5.93 & 2.98 & 7.47 & 4.01 & 12.21 & 6.67 & 5.70 & 2.99 \\ 
        & +\textsc{GraphAutoEnc} & 11.38 & 6.05 & 23.68 & 13.09 & 6.15 & 3.12 & 7.19 & 3.88 & 12.22 & 6.71 & 5.65 & 3.01 \\ \cmidrule{2-14}
        & +\textsc{NeighMean} & \textbf{11.81} & 6.20 & 22.43 & 12.07 & 8.21 & 4.21 & \textbf{12.40} & \textbf{6.62} & 12.54 & 6.74 & 8.54 & 4.45 \\
        & +\textsc{MultiHop} & \underline{11.79} & 6.13 & \underline{24.15} & \underline{13.28} & \underline{8.40} & \textbf{4.28} & \underline{12.35} & 6.59 & 12.63 & 6.80 & 8.53 & 4.49 \\
        & +\textsc{PersPageRank} & 11.41 & \underline{6.24} & 23.46 & 12.93 & \textbf{8.44} & \underline{4.26} & 12.33 & \underline{6.60} & \textbf{12.76} & \textbf{6.89} & \underline{8.55} & \underline{4.50} \\
        & +\textsc{Heat} & 11.63 & \textbf{6.25} & \textbf{24.60} & \textbf{13.41} & 8.26 & 4.21 & 12.17 & 6.48 & \underline{12.73} & \underline{6.88} & \textbf{8.59} & \textbf{4.51} \\
        \midrule
         \multirow{9}{*}{MMSSL} &  +\textsc{Zeros} &  13.40 &  \underline{11.36} &  28.83 &  21.42 &  8.84 &  6.55 &  \underline{12.06} &  8.65 &  \underline{13.76} &  10.05 &  9.11 &  7.00 \\
        &  +\textsc{Random} &  12.37 &  10.83 &  28.03 &  20.70 &  8.52 &  6.35 &  11.54 &  8.47 &  13.36 &  9.78 &  8.93 &  6.93 \\
        &  +\textsc{GlobalMean} &  13.07 &  11.00 &  28.89 &  21.55 &  8.86 &  6.58 &  12.00 &  8.68 &  13.58 &  9.87 &  9.21 &  7.09 \\
        \cmidrule{2-14}
        &  +\textsc{AutoEnc} &  \textbf{13.50} &  \textbf{11.38} &  28.85 &  21.47 &  8.83 &  6.58 &  11.73 &  8.55 &  13.61 &  9.98 &  9.05 &  7.01 \\        &  +\textsc{GraphAutoEnc} &  13.22 &  10.93 &  28.85 &  21.41 &  \textbf{8.95} &  \textbf{6.65} &  11.97 &  8.64 &  13.72 &  10.06 &  9.17 &  \underline{7.12} \\ \cmidrule{2-14}
        &  +\textsc{NeighMean} &  13.29 &  11.31 &  28.94 &  21.48 &  \underline{8.91} &  6.62 &  \underline{12.06} &  \underline{8.73} &  13.69 &  10.07 &  9.17 &  7.07 \\
        &  +\textsc{MultiHop} &  13.16 &  11.11 &  29.28 &  \textbf{21.92} &  8.90 &  \underline{6.63} &  12.05 &  \textbf{8.74} &  13.72 &  \underline{10.08} &  \underline{9.26} &  \textbf{7.14} \\
        &  +\textsc{PersPageRank} &  13.39 &  11.31 &  \underline{29.33} &  21.60 &  8.86 &  6.59 &  \textbf{12.13} &  \underline{8.73} &  13.72 &  10.01 &  9.22 &  7.10 \\
        &  +\textsc{Heat} &  \underline{13.48} &  11.29 &  \textbf{29.35} &  \underline{21.67} &  8.80 &  6.55 &  12.04 &  8.69 &  \textbf{13.77} &  \textbf{10.09} &  \textbf{9.27} &  \textbf{7.14}
        \\
        \midrule
         \multirow{9}{*}{DiffMM} &  +\textsc{Zeros} &  \textbf{13.34} &  \underline{7.36} &  27.83 &  15.97 &  8.69 &  4.52 &  11.59 &  6.23 &  13.04 &  7.03 &  9.23 &  \underline{4.89} \\
        &  +\textsc{Random} &  12.85 &  7.34 &  28.33 &  \textbf{16.61} &  8.65 &  \textbf{4.53} &  11.77 &  6.23 &  13.27 &  7.12 &  9.06 &  4.82 \\
        &  +\textsc{GlobalMean} &  \underline{13.20} &  \textbf{7.42} &  28.02 &  16.17 &  8.61 &  4.47 &  11.42 &  6.10 &  \textbf{13.35}&  \textbf{7.32}  &  8.92 &  4.68 \\
        \cmidrule{2-14}
        &  +\textsc{AutoEnc} &  13.17 &  7.31 &  28.04 &  16.17 &  8.71 &  4.51 &  11.65 &  6.19 &  13.25 &  7.16 &  9.26 &  4.88 \\ 
        &  +\textsc{GraphAutoEnc} &  12.79 &  7.19 &  28.37 &  16.27 &  8.67 &  4.50 &  \textbf{12.00} &  \textbf{6.37} &  13.08 &  7.03 &  9.18 &  4.82 \\ \cmidrule{2-14}
        &  +\textsc{NeighMean} &  12.89 &  7.04 &  \underline{28.50} &  16.31 &  8.68 &  \underline{4.52} &  11.77 &  6.25 &  \underline{13.31} &  \underline{7.19} &  9.25 &  \textbf{4.90} \\
        &  +\textsc{MultiHop} &  12.69 &  6.98 &  \textbf{28.73} &  16.38 &  \underline{8.72} &  4.50 &  11.82 &  6.30 &  13.24 &  7.12 &  \textbf{9.29} &  \textbf{4.90} \\
        &  +\textsc{PersPageRank} &  12.83 &  6.99 &  28.25 &  16.30 &  \textbf{8.81} &  4.49 &  \underline{11.93} &  \underline{6.34} &  13.25 &  7.11 &  9.26 &  4.88 \\
        &  +\textsc{Heat} &  12.60 &  6.93 &  28.49 &  \underline{16.46} &  8.66 &  4.50 &  11.92 &  6.33 &  13.27 &  7.13 &  \underline{9.28} &  \textbf{4.90} \\
        \bottomrule
    \end{tabular}
    \end{adjustbox}
\end{table*}

In the current analysis, we compare the recommendation performance of selected multimodal recommender systems in the dropped setting when equipped with traditional, autoencoder-, and graph-based imputation strategies. \Cref{tab:rq2} displays the results on all the recommendation datasets in terms of Recall@20 and nDCG@20. 

From a general overview, a consistent trend across all datasets and recommendation models suggests that graph-based imputation strategies (the ones proposed in this paper) settle as the best or second-best imputation approaches compared to all other techniques. However, it is important to highlight that for some recommendation models, such as VBPR, MGCN, and MMSSL, autoencoder-based imputations can still provide sufficient performance on specific datasets, maybe thanks to their ability to learn from the data how to recover one missing modality from the other available. 

Imputation-wise, we observe that, for traditional imputation, there is no recognisable trend, where each of the three approaches can provide improved performance depending on the considered recommendation model and dataset. Conversely, in terms of autoencoder-based imputation, we notice that, in most cases, \textsc{GraphAutoEnc} is the most successful solution among the two tested ones; this further confirms our intuition to adopt the item-item co-purchase graph structure to impute missing modalities. Finally, when it comes to graph-based imputation, approaches that propagate multimodal features at multiple hops (i.e., \textsc{MultiHop}), with more refined normalisation solutions (i.e., \textsc{PersPageRank} and \textsc{Heat}) tend to be the most powerful techniques.

In conclusion, our proposed training-free graph-based imputation methods largely outperform other families of imputation approaches on all the recommendation models and datasets. Moreover, we highlight the importance of leveraging the item-item co-purchase graph in the imputation, especially by exploring its topology at multiple hops. Finally, even against strong imputation strategies (such as the autoencoder-based ones which are trained to perform their task) our solutions can still provide high-quality imputations while being much more computationally-efficient (as they are training-free). 

\subsection{RQ3) Impact of sparsification rates and hop propagations}

\begin{figure}[!t]
\subfloat[VBPR]{
\begin{minipage}{\columnwidth}
\centering
    \input{figures/top_k_vbpr} \hspace{-1em}
    \input{figures/layers_vbpr}
    \vspace{-2em}
\end{minipage}
}
\\
\subfloat[MGCN]{
    \begin{minipage}{\columnwidth}
    \centering
    \input{figures/top_k_mgcn}  \hspace{-1em}
    \input{figures/layers_mgcn}
    \par \vspace{-1em} \centering \includegraphics[width=0.35\textwidth]{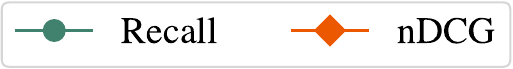}
    \end{minipage}
}
\caption{Performance variation for (a) VBPR and (b) MGCN with the \textsc{Heat} imputation method by considering different sparsification rates and propagation hops on the Sports dataset.}
\label{fig:sensitivity}
\end{figure}

This study serves to provide a hyper-parameter sensitivity analysis for the sparsification rate (TopN) and number of explored hops (Hops) in our graph-based imputation strategies. Spefically, we consider the performance variation on VBPR and MGCN equipped with the \textsc{Heat} imputation on the Sports recommendation dataset (\Cref{fig:sensitivity}).

Starting from VBPR, we see that the performance tends to slightly degrade when retaining bigger portions of neighbors for each node (i.e., higher TopN values). This aligns with the rationale that retaining only the most similar neighbors for each node should improve the informativeness level of the item-item co-purchase graph, thus improving the graph-based imputation and the recommendation performance. Regarding the Hops, we acknowledge that even with a few explorations, the graph-based imputation can reach a sufficiently high-quality reconstruction of missing features in VBPR, as we see improved recommendation performance for those settings and negligible performance improvements with increasing Hops. 

Trends are more or less aligned on MGCN in terms of Hops. However, focusing on the TopN, we clearly notice that retaining more neighbors for each node in the item-item co-purchase can, this time, improve the imputation and, consequently, the final recommendation performance. We ascribe this difference to the different nature of VBPR and MGCN, where the latter is a graph-based multimodal recommender system working on the user-item graph. In this respect, the careful reader would note that the item-item co-purchase graph used in our paper carries unexplicit information from the user-item graph (i.e., items that have been interacted by the same users). Thus, it seems reasonable that the \textsc{Heat} imputation working on a graph-based multimodal recommender system such as MGCN can benefit from larger neighbors to efficiently support the graph-based multimodal recommender system. 

All in all, the above investigation outlines that the proposed graph-based imputation require careful hyper-parameter selection depending on the strategies and methodologies adopted by the multimodal recommender system.

\subsection{RQ4) Impact of feature homophily}
\label{sec:rq4}

Inspired by~\cite{DBLP:conf/log/RossiK0C0B22}, we study the impact of feature homophily of the item-item co-purchase graph on the proposed graph-based imputation methods. To the best of our knowledge, this is the first time such an investigation is conducted in multimodal recommendation with missing modalities.

As the traditional and more common definition of node homophily~\cite{DBLP:conf/nips/ZhuYZHAK20} is not applicable to our setting, where no class nodes are available, we leverage the definition of feature homophily proposed in~\cite{DBLP:conf/sigir/Zhu0IKF24}. Specifically, we adapt its formulation to our case, where it is calculated for each modality in the system. 
%Specifically, let $\mathcal{E}^{\mathcal{I}} = \{(i, j) \in \mathcal{I} \times \mathcal{I} \; | \; \overline{\mathbf{R}}^{\mathcal{I}}_{ij} = 1\}$ be the set of item pairs that are connected with one edge in the item-item co-purchase graph (after it has been normalized and sparsified). Then, the feature homophily calculated on modality $m$ is obtained as: 
% \begin{equation}
% \footnotesize
%     H_m = \frac{\sum_{(i, j) \in \mathcal{E}^{\mathcal{I}}}  \mathbf{A}_{im} \; \cdot \; \mathbf{A}_{jm}}{\sqrt{\sum_{(i, j) \in \mathcal{E}^{\mathcal{I}}}  \mathbf{A}_{im} \; \cdot \;  \mathbf{A}_{im}} \; \cdot \; \sqrt{\sum_{(i, j) \in \mathcal{E}^{\mathcal{I}}} \mathbf{A}_{jm} \; \cdot \;  \mathbf{A}_{jm}}},
% \end{equation}
% where $ \mathbf{A}_{im} = (\mathbf{F}_{im}- \tilde{\mathbf{F}}_m)$, $ \mathbf{A}_{jm} = (\mathbf{F}_{jm}- \tilde{\mathbf{F}}_m)$, and $\tilde{\mathbf{F}}_m = \sum_{(i, j) \in \mathcal{E}^{\mathcal{I}}} (\mathbf{F}_{im} + \mathbf{F}_{jm}) / 2 |\mathcal{E}^{\mathcal{I}}|$. 
Note that we consider the feature homophily only in the \textit{dropped} setting as we need all available multimodal features to calculate it. The measure is high when close nodes in the graph tend to have similar features, and low otherwise.

\begin{figure}[!t]
\subfloat[VBPR]{
\begin{minipage}{\columnwidth}
\centering
    \input{figures/visual_vbpr} \hspace{-1em}
    \input{figures/textual_vbpr}
    \par \centering 
    \vspace{-1em}\includegraphics[width=0.45\textwidth]{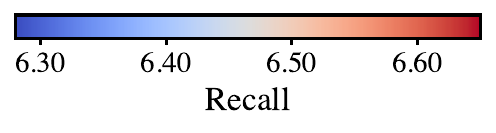}
\end{minipage}
}
\vspace{-1em}
\\
\subfloat[MGCN]{
    \begin{minipage}{\columnwidth}
    \centering
    \input{figures/visual_mgcn}
    \hspace{-1em}
    \input{figures/textual_mgcn}
    \par \centering \vspace{-1em} \includegraphics[width=0.45\textwidth]{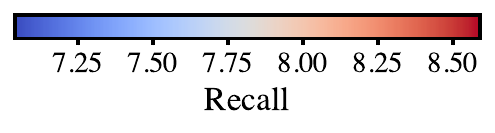}
    \end{minipage}
}
\caption{Performance variation (Recall) on Sports for (a) VBPR and (b) MGCN and with \textsc{Heat} imputation. Specifically, we display how the Recall changes with different levels of feature homophily ($x$-axis) and item-item sparsification ($y$-axis).}
\label{fig:rq4}
\end{figure}

\Cref{fig:rq4} depicts the performance variation for VBPR and MGCN on Sports dataset, equipped with \textsc{Heat} imputation, considering various levels of feature homophily ($x$-axis) and TopN sparsification ($y$-axis). The study is conducted on the Visual and Textual modalities separately, as we calculate the feature homophily on each of them.

The first two observable trends suggest that: (i) in general, feature homophily is very low and limited to the range [0, 0.20]; (ii) TopN sparsification and feature homophily are indirectly correspondent, as higher TopN values (so retaining more and low-quality neighbors of each node) inevitably lead to lower feature homophily and vice-versa. Indeed, observation (i) is confirmed by the outcomes from~\cite{DBLP:conf/sigir/Zhu0IKF24}, highlighting that the user-item graph in recommendation shows, in most cases, low levels of feature homophily.

Then, by considering the overall plots, we see again different trends between VBPR and MGCN. Regarding VBPR, we recognize higher recommendation performance coinciding with high levels of feature homophily, which seems to be aligned with what observed by the authors in~\cite{DBLP:conf/log/RossiK0C0B22}. Differently, on MGCN, we clearly notice higher recommendation performance for lower levels of homophily. Once again, this behavior might be ascribed to MGCN working with the user-item graph, whose perspective is strictly related to that of the item-item co-purchase graph we use for the graph-based imputations. Thus, as such imputations work by supporting the graph-based multimodal recommender systems in exploring vast portions of the user-item graph, they benefit from higher values of TopN and, consequently, lower levels of feature homophily. That is, feature homophily might be less impacting than for other multimodal recommender systems.

In conclusion, this study confirms the importance of hyper-parameter selection for the proposed graph-based approaches depending on the specific setting (as reported in the previous RQ). Moreover, regarding the generally low values of feature homophily, it underlines the necessity of better understanding and assessing this aspect, especially in the case this is preventing our solutions from providing even better performance. Thus, we will devote future work to the outlined direction.

\subsection{RQ5) Impact of missing features percentage for traditional, autoencoder-, and graph-based imputations}
\label{sec:rq5}

\begin{table*}[!t]
\caption{Performance variation (Recall) of FREEDOM and LGMRec (equipped with \textsc{GlobalMean}, \textsc{AutoEnc}, and \textsc{Heat} imputations) on the Baby dataset, with different percentages of missing features (i.e., Complete, 10\%, 50\%, and 90\%) on the Visual, Textual, and Visual + Textual modalities.}\label{tab:rq5}
    \centering
    \begin{adjustbox}{width=0.9\textwidth}
    \begin{tabular}{lcccclcccc}
    \toprule
        \multicolumn{5}{c}{\textbf{FREEDOM}} & \multicolumn{5}{c}{\textbf{LGMRec}} \\ \cmidrule(lr){1-5} \cmidrule(lr){6-10} 
        \multirow{2}{*}{\textbf{Imputation}} &  \multicolumn{4}{c}{Visual} & \multirow{2}{*}{\textbf{ Imputation}} & \multicolumn{4}{c}{ Visual} \\
        \cmidrule(lr){2-5} \cmidrule(lr){7-10}
        & Complete & 10\% & 50\% & 90\% & &  Complete &  10\% &  50\% &  90\% \\ \cmidrule(lr){1-10}
        +\textsc{GlobalMean} & 8.95 & 8.54$\pm$0.10 & %8.46$\pm$0.08 & 
        \textbf{8.60$\pm$0.09} 
        %& 8.47$\pm$0.05 
        & 8.47$\pm$0.09 &  +\textsc{GlobalMean} &  8.87 &  8.83$\pm$0.11 &  8.54$\pm$0.20 &  8.18$\pm$0.05 \\
        +\textsc{AutoEnc} & 8.95 & 8.60$\pm$0.15 
        %& 8.48$\pm$0.14 
        & 8.50$\pm$0.11 
        % & \textbf{8.52$\pm$0.14} 
        & \textbf{8.50$\pm$0.12} &  +\textsc{AutoEnc} &  8.87 &  8.85$\pm$0.16 &  8.76$\pm$0.16 &  \textbf{8.49$\pm$0.14} \\
        +\textsc{Heat} & 8.95 & \textbf{8.63$\pm$0.10} % & \textbf{8.52$\pm$0.09} 
        & 8.50$\pm$0.07 
        % & 8.46$\pm$0.12 
        & \textbf{8.50$\pm$0.05} &  +\textsc{Heat} &  8.87 &  \textbf{8.91$\pm$0.07} &  \textbf{8.82$\pm$0.06} &  8.48$\pm$0.04 \\
        \cmidrule{1-10}
        \multirow{2}{*}{\textbf{Imputation}} & \multicolumn{4}{c}{Textual} & \multirow{2}{*}{\textbf{ Imputation}} & \multicolumn{4}{c}{ Textual} \\
        \cmidrule(lr){2-5} \cmidrule(lr){7-10}
        & Complete & 10\% & 50\% & 90\% & &  Complete &  10\% &  50\% &  90\% \\ \cmidrule{1-10}
        +\textsc{GlobalMean} & 8.95 & 8.25$\pm$0.05 
        % & 7.73$\pm$0.12 
        & 7.46$\pm$0.20 
        % & 7.55$\pm$0.14 
        & 7.43$\pm$0.12 &  +\textsc{GlobalMean} &  8.87 &  8.74$\pm$0.12 &  8.19$\pm$0.15 &  7.59$\pm$0.17 \\
        +\textsc{AutoEnc} & 8.95 & \textbf{8.46$\pm$0.21} 
        % & \textbf{8.28$\pm$0.08} 
        & \textbf{7.92$\pm$0.08} 
        % & \textbf{7.72$\pm$0.12} 
        & \textbf{7.70$\pm$0.11} &  +\textsc{AutoEnc} &  8.87 &  8.81$\pm$0.07 &  8.25$\pm$0.18 &  7.40$\pm$0.05 \\
        +\textsc{Heat} & 8.95 & 8.33$\pm$0.11 
        % & 7.87$\pm$0.18 
        & 7.59$\pm$0.10 
        % & 7.36$\pm$0.07 
        & 7.19$\pm$0.17 &  +\textsc{Heat} &  8.87 &  \textbf{8.82$\pm$0.05} &  \textbf{8.36$\pm$0.14} &  \textbf{7.96$\pm$0.15} \\ \cmidrule{1-10}
        {\textbf{Imputation}} & \multicolumn{4}{c}{Visual + Textual} & {\textbf{ Imputation}} & \multicolumn{4}{c}{ Visual + Textual} \\
        \cmidrule(lr){2-5} \cmidrule(lr){7-10}
        & Complete & 10\% & 50\% & 90\% & &  Complete &  10\% &  50\% &  90\% \\ \cmidrule{1-10}
        +\textsc{GlobalMean} & 8.95 & 8.23$\pm$0.10 
        % & 7.58$\pm$0.08 
        & 7.24$\pm$0.18 
        % & 7.12$\pm$0.09 
        & 6.88$\pm$0.09 &  +\textsc{GlobalMean} &  8.87 &  8.52$\pm$0.14 &  7.49$\pm$0.21 &  \textbf{6.36$\pm$0.28} \\
        +\textsc{AutoEnc} & 8.95 & 7.16$\pm$0.15 
        % & 6.90$\pm$0.04 
        & 6.96$\pm$0.14 
        % & 6.76$\pm$0.11 
        & 6.88$\pm$0.13 &  +\textsc{AutoEnc} &  8.87 &  8.62$\pm$0.09 &  7.63$\pm$0.37 &  5.80$\pm$0.13 \\
        +\textsc{Heat} & 8.95 & \textbf{8.30$\pm$0.07} % & \textbf{7.80$\pm$0.11} 
        & \textbf{7.48$\pm$0.12} 
        % & \textbf{7.24$\pm$0.08} 
        & \textbf{6.99$\pm$0.08} &  +\textsc{Heat} &  8.87 &  \textbf{8.68$\pm$0.16} &  \textbf{7.83$\pm$0.21} &  6.24$\pm$0.14 \\
        \bottomrule
    \end{tabular}
    \end{adjustbox}
\end{table*}

This analysis is devoted to assessing the recommendation performance variation when considering selected imputation methods with different percentages of missing modalities. Specifically, we select FREEDOM and LGMRec as multimodal recommendation models, equipped with \textsc{GlobalMean}, \textsc{AutoEnc}, and \textsc{Heat} imputation methods, on the Baby recommendation dataset. Then, we simulate five scenarios where, either on the Visual, the Textual, or the Visual + Textual modalities, the 10\%, 50\%, or 90\% of items have missing features. For each of these configurations, we repeat the items sampling five times to ensure independence from the single-conducted sampling. Results are reported in \Cref{tab:rq5} in terms of Recall@20, where the first performance column refers to the Recall in the setting with all complete modalities, corresponding to our \textit{dropped} setting (where no item has missing multimodal features). 

Overall, results show that the graph-based imputation method (\textsc{Heat}) is the approach that performs the best in most cases; this becomes especially evident on LGMRec, where \textsc{Heat} is able to preserve recommendation performance even for higher percentages of missing modalities (e.g., see the Textual modality setting for 90\% of missing information, or the Visual + Textual modality setting for 50\% of missing information). The exception to the observed trend is the Textual modality for FREEDOM, where the autoencoder-based imputation (\textsc{AutoEnc}) is steadily the best. However, if we consider those scenarios with the percentage of missingness limited to 50\% (which are the most realistic ones according to \Cref{tab:datasets}, see the last column), our method settles as the best or second-to-best solution even in the Textual modality.

To summarise, graph-based imputation approaches demonstrate very high performance in most settings with various percentages of missing modalities, and in realistic scenarios, they still provide sufficiently improved results. Nevertheless, we will devote future directions of our work to delving into the results from the Textual missingness setting, which could represent an area for further investigation.

\subsection{RQ6) Impact of missing features on multimodal scenarios with more than two modalities}

We ultimately examine a multimodal recommendation scenario involving three modalities. In particular, we select the recent MicroLens dataset\footnote{\url{https://github.com/westlake-repl/MicroLens}.}~\cite{DBLP:conf/cikm/Ni0LFL0ZY25}. Unlike the Amazon datasets we used in the previous RQs, MicroLens comes with 3 modalities: visual, textual, and video. Starting from the original MicroLens-100K version, we apply an iterative $k$-core algorithm (with $k=7$) to retain users and items with the most interactions, resulting in the following statistics: 7,556 users, 4,628 items, and 81,505 interactions. Then, we simulate the missingness of modalities for 30\% of the items in the catalog.

\Cref{tab:rq6} reports results of MGCN when imputing missing modalities with two traditional strategies (i.e., \textsc{Random} and \textsc{GlobalMean}) and two graph-based techniques (i.e., \textsc{MultiHop} and \textsc{PersPageRank}). As previously observed in our analyses, this further confirms the efficacy of our proposed graph-based imputations, which can also effectively operate in multimodal scenarios with more than two modalities.

\begin{table}[!t]
\caption{Impact of traditional and graph-based imputation methods on the performance of MGCN for the MicroLens dataset having 3 modalities with 30\% of missing modalitie.}\label{tab:rq6}
    \centering
    \begin{tabular}{lcc}
    \toprule
         \textbf{Imputation} &  \textbf{Recall} &  \textbf{nDCG} \\ \cmidrule{1-3}
        +\textsc{Random} & 23.28$\pm$0.28 & 10.48$\pm$0.14 \\
        +\textsc{GlobalMean} & 23.38$\pm$0.35 & 10.55$\pm$0.18 \\
        +\textsc{MultiHop} & 23.41$\pm$0.37 & 10.45$\pm$0.22 \\
        +\textsc{PersPageRank} & \textbf{23.45$\pm$0.46} & \textbf{10.60$\pm$0.25} \\
        \bottomrule
    \end{tabular}
\end{table}

\section{Conclusion and future work}
\label{sec:conclusion}

In this paper, we tackled the problem of missing modalities in multimodal recommendation, which has found very limited attention in the recent literature. First, we formalised (for the first time) the issue by outlining differences with the problem of missing information in machine learning. Then, we proposed to leverage the item-item co-purchase graph to re-sketch the problem of missing modalities as a task of graph feature interpolation. Our solutions act as model-agnostic techniques applicable to any existing multimodal recommender systems without affecting their training (they run during the pre-processing phase), and are completely training-free. Extensive experiments on 7 datasets (with either 2 or 3 modalities), 14 (multimodal) recommenders, and against 5 imputation baselines (traditional and autoencoder-based imputation) outlined important aspects of our approaches. First, the graph-based imputations can preserve or even widen the performance gap between traditional and multimodal recommender systems. Second, the graph-based solutions largely outperform the other imputation strategies. Moreover, further analyses demonstrate that our approaches need careful hyperparameter selection to adapt to the specific multimodal recommender system. Despite their largely-demonstrated efficacy, final results regarding the impact of feature homophily and the percentage of missing features on each modality underline some performance limitations in our approaches, paving the way for interesting future directions of the current work. Among others, we plan to re-design our graph-based imputation strategies with more refined sparsifications~\cite{DBLP:conf/kdd/Xv0GGLDXZ23} and integrate them into the end-to-end recommendation pipeline with the multimodal recommendation model. This would ensure that the framework exploits the information-rich item-item similarity graph as done in the related literature~\cite{zhou2023enhancing, DBLP:conf/aaai/00030000N25, DBLP:conf/mm/ZhouS23}, along with ad-hoc modality alignment constraints to provide further guidance signal to the imputation of missing modalities~\cite{DBLP:conf/aaai/00030000N25, DBLP:conf/www/WeiHXZ23, DBLP:journals/tmm/TaoLXWYHC23, DBLP:conf/www/ZhouZLZMWYJ23}, and works effectively in cold-start/noisy scenarios.

\section*{Acknowledgments}
\noindent D.M. and F.D.M. acknowledge the support of the Innov4-ePiK project managed by the French National Research Agency under the 4th PIA, integrated into France2030 (ANR-23-RHUS-0002). C.P. and T.D.N. acknowledge the support of the P+ARTS project funded by the European Union Next-GenerationEU (NRRP – M4C1, Investment 3.4, INTAFAM00037, CUP: G43C24000640006). The authors acknowledge Inria, Mesocentre, and the CINECA (ISCRA initiative) for high-performance computing resources.

\bibliographystyle{IEEEtran}
\bibliography{references}

@inproceedings{DBLP:conf/icml/KondorL02,
  author       = {Risi Kondor and
                  John D. Lafferty},
  title        = {Diffusion Kernels on Graphs and Other Discrete Input Spaces},
  booktitle    = {{ICML}},
  year         = {2002}
}

@inproceedings{DBLP:conf/cvpr/LeeTCL23,
  author       = {Yi{-}Lun Lee and
                  Yi{-}Hsuan Tsai and
                  Wei{-}Chen Chiu and
                  Chen{-}Yu Lee},
  title        = {Multimodal Prompting with Missing Modalities for Visual Recognition},
  booktitle    = {{CVPR}},
  publisher    = {{IEEE}},
  year         = {2023}
}

@inproceedings{DBLP:conf/www/WeiTXJH24,
  author       = {Wei Wei and
                  Jiabin Tang and
                  Lianghao Xia and
                  Yangqin Jiang and
                  Chao Huang},
  title        = {PromptMM: Multi-Modal Knowledge Distillation for Recommendation with
                  Prompt-Tuning},
  booktitle    = {{WWW}},
  year         = {2024}
}

@inproceedings{DBLP:conf/apweb/WangYCJYKWHL24,
  author       = {Jie Wang and
                  Fajie Yuan and
                  Mingyue Cheng and
                  Joemon M. Jose and
                  Chenyun Yu and
                  Beibei Kong and
                  Zhijin Wang and
                  Bo Hu and
                  Zang Li},
  title        = {TransRec: Learning Transferable Recommendation from Mixture-of-Modality
                  Feedback},
  booktitle    = {APWeb/WAIM {(2)}},
  year         = {2024}
}

@article{DBLP:journals/inffus/PanPCC26,
  author       = {Lin Pan and
                  Zhiqiang Pan and
                  Fei Cai and
                  Honghui Chen},
  title        = {Multimodal recommender systems: {A} survey of representation, modeling,
                  and optimization},
  journal      = {Inf. Fusion},
  volume       = {128},
  pages        = {103991},
  year         = {2026}
}

@inproceedings{DBLP:conf/cikm/Ni0LFL0ZY25,
  author       = {Yongxin Ni and
                  Yu Cheng and
                  Xiangyan Liu and
                  Junchen Fu and
                  Youhua Li and
                  Xiangnan He and
                  Yongfeng Zhang and
                  Fajie Yuan},
  title        = {A Content-Driven Micro-Video Recommendation Dataset at Scale},
  booktitle    = {{CIKM}},
  year         = {2025}
}

@article{DBLP:journals/corr/abs-2502-15711,
  author       = {Jinfeng Xu and
                  Zheyu Chen and
                  Shuo Yang and
                  Jinze Li and
                  Wei Wang and
                  Xiping Hu and
                  Steven Hoi and
                  Edith C. H. Ngai},
  title        = {A Survey on Multimodal Recommender Systems: Recent Advances and Future
                  Directions},
  journal      = {CoRR},
  volume       = {abs/2502.15711},
  year         = {2025}
}

@article{DBLP:journals/jbd/EmmanuelMMSMT21,
  author       = {Tlamelo Emmanuel and
                  Thabiso M. Maupong and
                  Dimane Mpoeleng and
                  Thabo Semong and
                  Banyatsang Mphago and
                  Oteng Tabona},
  title        = {A survey on missing data in machine learning},
  journal      = {J. Big Data},
  volume       = {8},
  number       = {1},
  pages        = {140},
  year         = {2021}
}

@techreport{ilprints422,
          number = {1999-66},
           month = {November},
          author = {Lawrence Page and Sergey Brin and Rajeev Motwani and Terry Winograd},
           title = {The PageRank Citation Ranking: Bringing Order to the Web.},
            type = {Technical Report},
       publisher = {Stanford InfoLab},
            year = {1999},
     institution = {Stanford InfoLab},
    }

@inproceedings{DBLP:conf/nips/KlicperaWG19,
  author       = {Johannes Klicpera and
                  Stefan Wei{\ss}enberger and
                  Stephan G{\"{u}}nnemann},
  title        = {Diffusion Improves Graph Learning},
  booktitle    = {NeurIPS},
  year         = {2019}
}

@inproceedings{DBLP:conf/mm/JiangX0LLH24,
  author       = {Yangqin Jiang and
                  Lianghao Xia and
                  Wei Wei and
                  Da Luo and
                  Kangyi Lin and
                  Chao Huang},
  title        = {DiffMM: Multi-Modal Diffusion Model for Recommendation},
  booktitle    = {{ACM} Multimedia},
  year         = {2024}
}

@inproceedings{DBLP:conf/mm/WeiWN0HC19,
  author    = {Yinwei Wei and
               Xiang Wang and
               Liqiang Nie and
               Xiangnan He and
               Richang Hong and
               Tat{-}Seng Chua},
  title     = {{MMGCN:} Multi-modal Graph Convolution Network for Personalized Recommendation
               of Micro-video},
  booktitle = {{ACM} Multimedia},
  year      = {2019}
}

@article{DBLP:journals/tmm/WangWYWSN23,
  author       = {Qifan Wang and
                  Yinwei Wei and
                  Jianhua Yin and
                  Jianlong Wu and
                  Xuemeng Song and
                  Liqiang Nie},
  title        = {DualGNN: Dual Graph Neural Network for Multimedia Recommendation},
  journal      = {{IEEE} Trans. Multim.},
  volume       = {25},
  pages        = {1074--1084},
  year         = {2023}
}

@inproceedings{DBLP:conf/iclr/KipfW17,
  author    = {Thomas N. Kipf and
               Max Welling},
  title     = {Semi-Supervised Classification with Graph Convolutional Networks},
  booktitle = {{ICLR} (Poster)},
  publisher = {OpenReview.net},
  year      = {2017}
}

@inproceedings{DBLP:conf/sigir/Wang0WFC19,
  author    = {Xiang Wang and
               Xiangnan He and
               Meng Wang and
               Fuli Feng and
               Tat{-}Seng Chua},
  title     = {Neural Graph Collaborative Filtering},
  booktitle = {{SIGIR}},
  year      = {2019}
}

@incollection{DBLP:books/acm/18/BaltrusaitisAM18,
  author    = {Tadas Baltrusaitis and
               Chaitanya Ahuja and
               Louis{-}Philippe Morency},
  title     = {Challenges and applications in multimodal machine learning},
  booktitle = {The Handbook of Multimodal-Multisensor Interfaces, Volume 2 {(2)}},
  year      = {2018}
}

@inproceedings{DBLP:conf/sigir/0001DWLZ020,
  author    = {Xiangnan He and
               Kuan Deng and
               Xiang Wang and
               Yan Li and
               Yong{-}Dong Zhang and
               Meng Wang},
  title     = {LightGCN: Simplifying and Powering Graph Convolution Network for Recommendation},
  booktitle = {{SIGIR}},
  year      = {2020}
}

@inproceedings{DBLP:conf/cvpr/HeZRS16,
  author    = {Kaiming He and
               Xiangyu Zhang and
               Shaoqing Ren and
               Jian Sun},
  title     = {Deep Residual Learning for Image Recognition},
  booktitle = {{CVPR}},
  year      = {2016}
}

@inproceedings{DBLP:conf/uai/RendleFGS09,
  author    = {Steffen Rendle and
               Christoph Freudenthaler and
               Zeno Gantner and
               Lars Schmidt{-}Thieme},
  title     = {{BPR:} Bayesian Personalized Ranking from Implicit Feedback},
  booktitle = {{UAI}},
  year      = {2009}
}

@article{DBLP:journals/tmm/MinJJ20,
  author    = {Weiqing Min and
               Shuqiang Jiang and
               Ramesh C. Jain},
  title     = {Food Recommendation: Framework, Existing Solutions, and Challenges},
  journal   = {{IEEE} Trans. Multim.},
  year      = {2020}
}

@inproceedings{DBLP:conf/aaai/00030000N25,
  author       = {Jinfeng Xu and
                  Zheyu Chen and
                  Shuo Yang and
                  Jinze Li and
                  Hewei Wang and
                  Edith C. H. Ngai},
  title        = {{MENTOR:} Multi-level Self-supervised Learning for Multimodal Recommendation},
  booktitle    = {{AAAI}},
  year         = {2025}
}

@article{DBLP:journals/siamrev/Strawderman89,
  author       = {William E. Strawderman},
  title        = {Statistical Analysis with Missing Data (Roderick J. A. Little and
                  Donald B. Rubin)},
  journal      = {{SIAM} Rev.},
  volume       = {31},
  number       = {2},
  pages        = {348--349},
  year         = {1989}
}

@inproceedings{DBLP:conf/mm/LiuYLWTZSM21,
  author    = {Yong Liu and
               Susen Yang and
               Chenyi Lei and
               Guoxin Wang and
               Haihong Tang and
               Juyong Zhang and
               Aixin Sun and
               Chunyan Miao},
  title     = {Pre-training Graph Transformer with Multimodal Side Information for
               Recommendation},
  booktitle = {{ACM} Multimedia},
  year      = {2021}
}

@inproceedings{DBLP:conf/mmasia/Zhou23,
  author       = {Xin Zhou},
  title        = {MMRec: Simplifying Multimodal Recommendation},
  booktitle    = {MMAsia (Workshops)},
  year         = {2023}
}

@article{DBLP:journals/tors/MalitestaCPMNS25,
  author       = {Daniele Malitesta and
                  Giandomenico Cornacchia and
                  Claudio Pomo and
                  Felice Antonio Merra and
                  Tommaso Di Noia and
                  Eugenio Di Sciascio},
  title        = {Formalizing Multimedia Recommendation through Multimodal Deep Learning},
  journal      = {Trans. Recomm. Syst.},
  volume       = {3},
  number       = {3},
  pages        = {37:1--37:33},
  year         = {2025}
}

@inproceedings{DBLP:conf/mm/Zhang00WWW21,
  author    = {Jinghao Zhang and
               Yanqiao Zhu and
               Qiang Liu and
               Shu Wu and
               Shuhui Wang and
               Liang Wang},
  title     = {Mining Latent Structures for Multimedia Recommendation},
  booktitle = {{ACM} Multimedia},
  year      = {2021}
}

@inproceedings{DBLP:conf/nips/ZhuYZHAK20,
  author       = {Jiong Zhu and
                  Yujun Yan and
                  Lingxiao Zhao and
                  Mark Heimann and
                  Leman Akoglu and
                  Danai Koutra},
  title        = {Beyond Homophily in Graph Neural Networks: Current Limitations and
                  Effective Designs},
  booktitle    = {NeurIPS},
  year         = {2020}
}

@inproceedings{DBLP:conf/sigir/BaiWHCHZHW24,
  author       = {Haoyue Bai and
                  Le Wu and
                  Min Hou and
                  Miaomiao Cai and
                  Zhuangzhuang He and
                  Yuyang Zhou and
                  Richang Hong and
                  Meng Wang},
  title        = {Multimodality Invariant Learning for Multimedia-Based New Item Recommendation},
  booktitle    = {{SIGIR}},
  year         = {2024}
}

@inproceedings{DBLP:conf/recsys/GanhorMHNS24,
  author       = {Christian Ganh{\"{o}}r and
                  Marta Moscati and
                  Anna Hausberger and
                  Shah Nawaz and
                  Markus Schedl},
  title        = {A Multimodal Single-Branch Embedding Network for Recommendation in
                  Cold-Start and Missing Modality Scenarios},
  booktitle    = {RecSys},
  year         = {2024}
}

@inproceedings{DBLP:conf/mm/LinTZLW0WY23,
  author       = {Zhenghong Lin and
                  Yanchao Tan and
                  Yunfei Zhan and
                  Weiming Liu and
                  Fan Wang and
                  Chaochao Chen and
                  Shiping Wang and
                  Carl Yang},
  title        = {Contrastive Intra- and Inter-Modality Generation for Enhancing Incomplete
                  Multimedia Recommendation},
  booktitle    = {{ACM} Multimedia},
  year         = {2023}
}

@inproceedings{DBLP:conf/log/RossiK0C0B22,
  author       = {Emanuele Rossi and
                  Henry Kenlay and
                  Maria I. Gorinova and
                  Benjamin Paul Chamberlain and
                  Xiaowen Dong and
                  Michael M. Bronstein},
  title        = {On the Unreasonable Effectiveness of Feature Propagation in Learning
                  on Graphs With Missing Node Features},
  booktitle    = {LoG},
  year         = {2022}
}

@inproceedings{DBLP:conf/kdd/Steck10,
  author       = {Harald Steck},
  title        = {Training and testing of recommender systems on data missing not at
                  random},
  booktitle    = {{KDD}},
  year         = {2010}
}

@article{DBLP:journals/tkde/LiuCZLN22,
  author       = {Fan Liu and
                  Zhiyong Cheng and
                  Lei Zhu and
                  Chenghao Liu and
                  Liqiang Nie},
  title        = {An Attribute-Aware Attentive {GCN} Model for Attribute Missing in
                  Recommendation},
  journal      = {{IEEE} Trans. Knowl. Data Eng.},
  volume       = {34},
  number       = {9},
  pages        = {4077--4088},
  year         = {2022}
}

@inproceedings{DBLP:conf/cikm/ShiZYZHLM19,
  author       = {Shaoyun Shi and
                  Min Zhang and
                  Xinxing Yu and
                  Yongfeng Zhang and
                  Bin Hao and
                  Yiqun Liu and
                  Shaoping Ma},
  title        = {Adaptive Feature Sampling for Recommendation with Missing Content
                  Feature Values},
  booktitle    = {{CIKM}},
  year         = {2019}
}

@inproceedings{DBLP:conf/emnlp/WangNL18,
  author       = {Cheng Wang and
                  Mathias Niepert and
                  Hui Li},
  title        = {{LRMM:} Learning to Recommend with Missing Modalities},
  booktitle    = {{EMNLP}},
  year         = {2018}
}

@inproceedings{DBLP:conf/dsaa/VernadeC15,
  author       = {Claire Vernade and
                  Olivier Capp{\'{e}}},
  title        = {Learning from missing data using selection bias in movie recommendation},
  booktitle    = {{DSAA}},
  year         = {2015}
}

@inproceedings{DBLP:conf/nss/Takasu11,
  author       = {Atsuhiro Takasu},
  title        = {A multicriteria recommendation method for data with missing rating
                  scores},
  booktitle    = {{ICDKE}},
  year         = {2011}
}

@inproceedings{DBLP:conf/icdm/Chen0ESFC18,
  author       = {Jiawei Chen and
                  Can Wang and
                  Martin Ester and
                  Qihao Shi and
                  Yan Feng and
                  Chun Chen},
  title        = {Social Recommendation with Missing Not at Random Data},
  booktitle    = {{ICDM}},
  year         = {2018}
}

@article{DBLP:journals/tkde/ZhengWXLW22,
  author       = {Xiaolin Zheng and
                  Menghan Wang and
                  Renjun Xu and
                  Jianmeng Li and
                  Yan Wang},
  title        = {Modeling Dynamic Missingness of Implicit Feedback for Sequential Recommendation},
  journal      = {{IEEE} Trans. Knowl. Data Eng.},
  volume       = {34},
  number       = {1},
  pages        = {405--418},
  year         = {2022}
}

@inproceedings{DBLP:conf/nips/WangGZZ18,
  author       = {Menghan Wang and
                  Mingming Gong and
                  Xiaolin Zheng and
                  Kun Zhang},
  title        = {Modeling Dynamic Missingness of Implicit Feedback for Recommendation},
  booktitle    = {NeurIPS},
  year         = {2018}
}

@inproceedings{DBLP:conf/cikm/MalitestaRPNM24,
  author       = {Daniele Malitesta and
                  Emanuele Rossi and
                  Claudio Pomo and
                  Tommaso Di Noia and
                  Fragkiskos D. Malliaros},
  title        = {Do We Really Need to Drop Items with Missing Modalities in Multimodal
                  Recommendation?},
  booktitle    = {{CIKM}},
  year         = {2024}
}

@inproceedings{DBLP:conf/recsys/OramasNSS17,
  author    = {Sergio Oramas and
               Oriol Nieto and
               Mohamed Sordo and
               Xavier Serra},
  title     = {A Deep Multimodal Approach for Cold-start Music Recommendation},
  booktitle = {DLRS@RecSys},
  year      = {2017}
}

@inproceedings{DBLP:conf/sigir/WuWF0CLX21,
  author       = {Jiancan Wu and
                  Xiang Wang and
                  Fuli Feng and
                  Xiangnan He and
                  Liang Chen and
                  Jianxun Lian and
                  Xing Xie},
  title        = {Self-supervised Graph Learning for Recommendation},
  booktitle    = {{SIGIR}},
  year         = {2021}
}

@inproceedings{DBLP:conf/aaai/GuoL0WSR24,
  author       = {Zhiqiang Guo and
                  Jianjun Li and
                  Guohui Li and
                  Chaoyang Wang and
                  Si Shi and
                  Bin Ruan},
  title        = {LGMRec: Local and Global Graph Learning for Multimodal Recommendation},
  booktitle    = {{AAAI}},
  year         = {2024}
}

@inproceedings{DBLP:conf/mm/Yu0LB23,
  author       = {Penghang Yu and
                  Zhiyi Tan and
                  Guanming Lu and
                  Bing{-}Kun Bao},
  title        = {Multi-View Graph Convolutional Network for Multimedia Recommendation},
  booktitle    = {{ACM} Multimedia},
  year         = {2023}
}

@inproceedings{DBLP:conf/wsdm/OngK25,
  author       = {Rongqing Kenneth Ong and
                  Andy W. H. Khong},
  title        = {Spectrum-based Modality Representation Fusion Graph Convolutional
                  Network for Multimodal Recommendation},
  booktitle    = {{WSDM}},
  year         = {2025}
}

@inproceedings{DBLP:conf/ecai/Zhou0Z023,
  author       = {Hongyu Zhou and
                  Xin Zhou and
                  Lingzi Zhang and
                  Zhiqi Shen},
  title        = {Enhancing Dyadic Relations with Homogeneous Graphs for Multimodal
                  Recommendation},
  booktitle    = {{ECAI}},
  year         = {2023}
}

@inproceedings{DBLP:conf/kdd/Xv0GGLDXZ23,
  author       = {Guipeng Xv and
                  Chen Lin and
                  Wanxian Guan and
                  Jinping Gou and
                  Xubin Li and
                  Hongbo Deng and
                  Jian Xu and
                  Bo Zheng},
  title        = {E-commerce Search via Content Collaborative Graph Neural Network},
  booktitle    = {{KDD}},
  year         = {2023}
}

@article{DBLP:journals/tmm/YiC24,
  author       = {Jing Yi and
                  Zhenzhong Chen},
  title        = {Variational Mixture of Stochastic Experts Auto-Encoder for Multi-Modal
                  Recommendation},
  journal      = {{IEEE} Trans. Multim.},
  volume       = {26},
  pages        = {8941--8954},
  year         = {2024}
}

@article{DBLP:journals/tmm/ZhouM24,
  author       = {Xin Zhou and
                  Chunyan Miao},
  title        = {Disentangled Graph Variational Auto-Encoder for Multimodal Recommendation
                  With Interpretability},
  journal      = {{IEEE} Trans. Multim.},
  volume       = {26},
  pages        = {7543--7554},
  year         = {2024}
}

@inproceedings{DBLP:conf/mm/Su0L0024,
  author       = {Hongzu Su and
                  Jingjing Li and
                  Fengling Li and
                  Ke Lu and
                  Lei Zhu},
  title        = {{SOIL:} Contrastive Second-Order Interest Learning for Multimodal
                  Recommendation},
  booktitle    = {{ACM} Multimedia},
  year         = {2024}
}

@article{DBLP:journals/tmm/TaoLXWYHC23,
  author       = {Zhulin Tao and
                  Xiaohao Liu and
                  Yewei Xia and
                  Xiang Wang and
                  Lifang Yang and
                  Xianglin Huang and
                  Tat{-}Seng Chua},
  title        = {Self-Supervised Learning for Multimedia Recommendation},
  journal      = {{IEEE} Trans. Multim.},
  volume       = {25},
  pages        = {5107--5116},
  year         = {2023}
}

@inproceedings{DBLP:conf/www/WeiHXZ23,
  author       = {Wei Wei and
                  Chao Huang and
                  Lianghao Xia and
                  Chuxu Zhang},
  title        = {Multi-Modal Self-Supervised Learning for Recommendation},
  booktitle    = {{WWW}},
  year         = {2023}
}

@inproceedings{DBLP:conf/cikm/LinMWLZ024,
  author       = {Guojiao Lin and
                  Zhen Meng and
                  Dongjie Wang and
                  Qingqing Long and
                  Yuanchun Zhou and
                  Meng Xiao},
  title        = {{GUME:} Graphs and User Modalities Enhancement for Long-Tail Multimodal
                  Recommendation},
  booktitle    = {{CIKM}},
  year         = {2024}
}

@inproceedings{DBLP:conf/www/ZhouZLZMWYJ23,
  author       = {Xin Zhou and
                  Hongyu Zhou and
                  Yong Liu and
                  Zhiwei Zeng and
                  Chunyan Miao and
                  Pengwei Wang and
                  Yuan You and
                  Feijun Jiang},
  title        = {Bootstrap Latent Representations for Multi-modal Recommendation},
  booktitle    = {{WWW}},
  year         = {2023}
}

@inproceedings{DBLP:conf/mm/WeiWN0C20,
  author    = {Yinwei Wei and
               Xiang Wang and
               Liqiang Nie and
               Xiangnan He and
               Tat{-}Seng Chua},
  title     = {Graph-Refined Convolutional Network for Multimedia Recommendation
               with Implicit Feedback},
  booktitle = {{ACM} Multimedia},
  year      = {2020}
}

@article{DBLP:journals/tmm/CaiQFX22,
  author    = {Desheng Cai and
               Shengsheng Qian and
               Quan Fang and
               Changsheng Xu},
  title     = {Heterogeneous Hierarchical Feature Aggregation Network for Personalized
               Micro-Video Recommendation},
  journal   = {{IEEE} Trans. Multim.},
  volume    = {24},
  pages     = {805--818},
  year      = {2022}
}

@inproceedings{DBLP:conf/sigir/ChenCXZ0QZ19,
  author    = {Xu Chen and
               Hanxiong Chen and
               Hongteng Xu and
               Yongfeng Zhang and
               Yixin Cao and
               Zheng Qin and
               Hongyuan Zha},
  title     = {Personalized Fashion Recommendation with Visual Explanations based
               on Multimodal Attention Network: Towards Visually Explainable Recommendation},
  booktitle = {{SIGIR}},
  year      = {2019}
}

@article{DBLP:journals/tmm/YiC22,
  author    = {Jing Yi and
               Zhenzhong Chen},
  title     = {Multi-Modal Variational Graph Auto-Encoder for Recommendation Systems},
  journal   = {{IEEE} Trans. Multim.},
  year      = {2022}
}

@inproceedings{DBLP:conf/sigir/Yi0OM22,
  author    = {Zixuan Yi and
               Xi Wang and
               Iadh Ounis and
               Craig MacDonald},
  title     = {Multi-modal Graph Contrastive Learning for Micro-video Recommendation},
  booktitle = {{SIGIR}},
  year      = {2022}
}

@inproceedings{DBLP:conf/mm/ZhouS23,
  author       = {Xin Zhou and
                  Zhiqi Shen},
  title        = {A Tale of Two Graphs: Freezing and Denoising Graph Structures for
                  Multimodal Recommendation},
  booktitle    = {{ACM} Multimedia},
  year         = {2023}
}

@inproceedings{DBLP:conf/kdd/ZhangCMZWWZ22,
  author    = {Chaohe Zhang and
               Xu Chu and
               Liantao Ma and
               Yinghao Zhu and
               Yasha Wang and
               Jiangtao Wang and
               Junfeng Zhao},
  title     = {M3Care: Learning with Missing Modalities in Multimodal Healthcare
               Data},
  booktitle = {{KDD}},
  year      = {2022}
}

@inproceedings{DBLP:conf/aaai/MaRZTWP21,
  author    = {Mengmeng Ma and
               Jian Ren and
               Long Zhao and
               Sergey Tulyakov and
               Cathy Wu and
               Xi Peng},
  title     = {{SMIL:} Multimodal Learning with Severely Missing Modality},
  booktitle = {{AAAI}},
  year      = {2021}
}

@inproceedings{DBLP:conf/emnlp/ReimersG19,
  author       = {Nils Reimers and
                  Iryna Gurevych},
  title        = {Sentence-BERT: Sentence Embeddings using Siamese BERT-Networks},
  booktitle    = {{EMNLP/IJCNLP} {(1)}},
  year         = {2019}
}

@inproceedings{DBLP:conf/cikm/KimLSK22,
  author       = {Taeri Kim and
                  Yeon{-}Chang Lee and
                  Kijung Shin and
                  Sang{-}Wook Kim},
  title        = {{MARIO:} Modality-Aware Attention and Modality-Preserving Decoders
                  for Multimedia Recommendation},
  booktitle    = {{CIKM}},
  year         = {2022}
}

@inproceedings{DBLP:conf/aaai/Yu0LB25,
  author       = {Penghang Yu and
                  Zhiyi Tan and
                  Guanming Lu and
                  Bing{-}Kun Bao},
  title        = {Mind Individual Information! Principal Graph Learning for Multimedia
                  Recommendation},
  booktitle    = {{AAAI}},
  year         = {2025}
}

@article{DBLP:journals/corr/abs-2302-04473,
  author       = {Hongyu Zhou and
                  Xin Zhou and
                  Zhiwei Zeng and
                  Lingzi Zhang and
                  Zhiqi Shen},
  title        = {A Comprehensive Survey on Multimodal Recommender Systems: Taxonomy,
                  Evaluation, and Future Directions},
  journal      = {CoRR},
  volume       = {abs/2302.04473},
  year         = {2023}
}

@inproceedings{DBLP:conf/aaai/HeM16,
  author    = {Ruining He and
               Julian J. McAuley},
  title     = {{VBPR:} Visual Bayesian Personalized Ranking from Implicit Feedback},
  booktitle = {{AAAI}},
  year      = {2016}
}

@inproceedings{DBLP:conf/wsdm/WeiRTWSCWYH24,
  author       = {Wei Wei and
                  Xubin Ren and
                  Jiabin Tang and
                  Qinyong Wang and
                  Lixin Su and
                  Suqi Cheng and
                  Junfeng Wang and
                  Dawei Yin and
                  Chao Huang},
  title        = {LLMRec: Large Language Models with Graph Augmentation for Recommendation},
  booktitle    = {{WSDM}},
  year         = {2024}
}

@inproceedings{DBLP:conf/kdd/LiuZYDD0ZZD24,
  author       = {Qijiong Liu and
                  Jieming Zhu and
                  Yanting Yang and
                  Quanyu Dai and
                  Zhaocheng Du and
                  Xiao{-}Ming Wu and
                  Zhou Zhao and
                  Rui Zhang and
                  Zhenhua Dong},
  title        = {Multimodal Pretraining, Adaptation, and Generation for Recommendation:
                  {A} Survey},
  booktitle    = {{KDD}},
  year         = {2024}
}

@article{DBLP:journals/csur/LiuHXZGWLT25,
  author       = {Qidong Liu and
                  Jiaxi Hu and
                  Yutian Xiao and
                  Xiangyu Zhao and
                  Jingtong Gao and
                  Wanyu Wang and
                  Qing Li and
                  Jiliang Tang},
  title        = {Multimodal Recommender Systems: {A} Survey},
  journal      = {{ACM} Comput. Surv.},
  volume       = {57},
  number       = {2},
  pages        = {26:1--26:17},
  year         = {2025}
}

@article{zhou2023enhancing,
  title={Enhancing Dyadic Relations with Homogeneous Graphs for Multimodal Recommendation},
  author={Zhou, Hongyu and Zhou, Xin and Shen, Zhiqi},
  journal={ECAI},
  year={2023}
}

@inproceedings{DBLP:conf/sigir/Zhu0IKF24,
  author       = {Jing Zhu and
                  Xiang Song and
                  Vassilis N. Ioannidis and
                  Danai Koutra and
                  Christos Faloutsos},
  title        = {TouchUp-G: Improving Feature Representation through Graph-Centric
                  Finetuning},
  booktitle    = {{SIGIR}},
  year         = {2024}
}

%\newpage

\section{Biography Section}

%\vspace{-3.5em}

\begin{IEEEbiography}[{\includegraphics[width=1in,height=1.25in,clip,keepaspectratio]{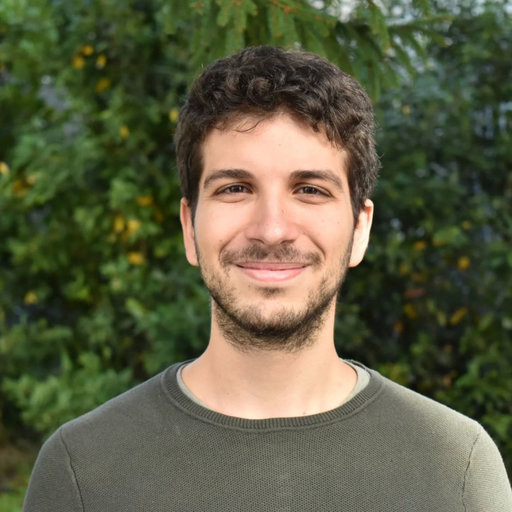}}]{Daniele Malitesta}
is a Postdoctoral researcher at Université Paris-Saclay. Previously, he pursued a PhD at Politecnico di Bari (Italy), where he focused on graph-based multimodal recommendation. His works appear in venues such as ECML-PKDD, The Web Conference, SIGIR, and RecSys. Recently, he has given lectures and tutorials on his past and current research topics, which mainly focus on fairness in graph generative models and missing multimodal information in graph machine learning.
\end{IEEEbiography}

%\vspace{-4em}

\begin{IEEEbiography}[{\includegraphics[width=1in,height=1.25in,clip,keepaspectratio]{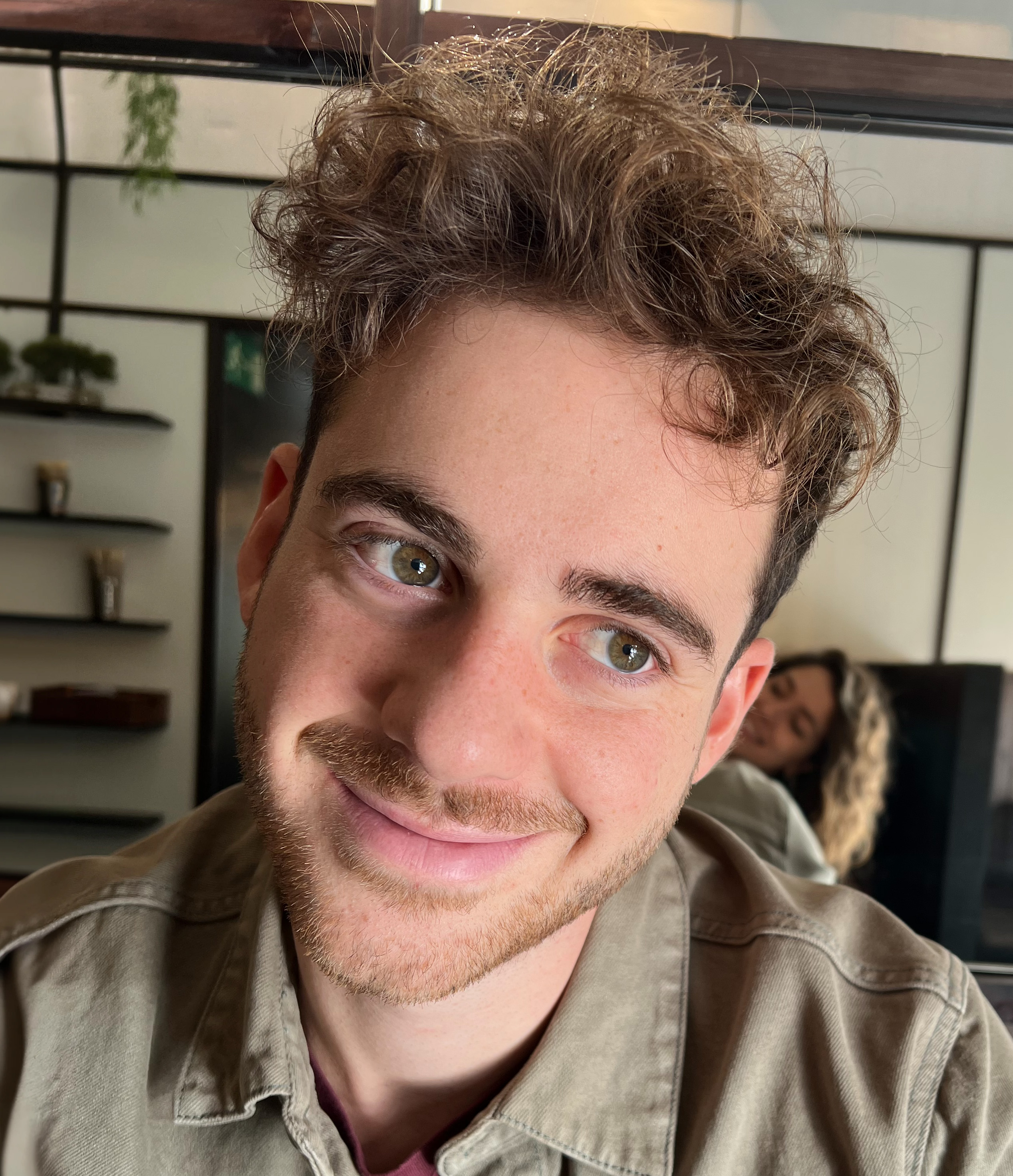}}]{Emanuele Rossi} is a Machine Learning Researcher at Vant AI, where he develops generative models at the intersection of machine learning and biology. He holds a Ph.D. from Imperial College London, where he was supervised by Prof. Michael Bronstein and focused on Graph Neural Networks.
His research appears at NeurIPS, ICML, ICLR, AAAI, and RecSys. Additionally, he has co-organized the Temporal Graph Learning Workshop at NeurIPS in both 2023 and 2024. 
\end{IEEEbiography}

%\vspace{-4em}

\begin{IEEEbiography}[{\includegraphics[width=1in,height=1.25in,clip,keepaspectratio]{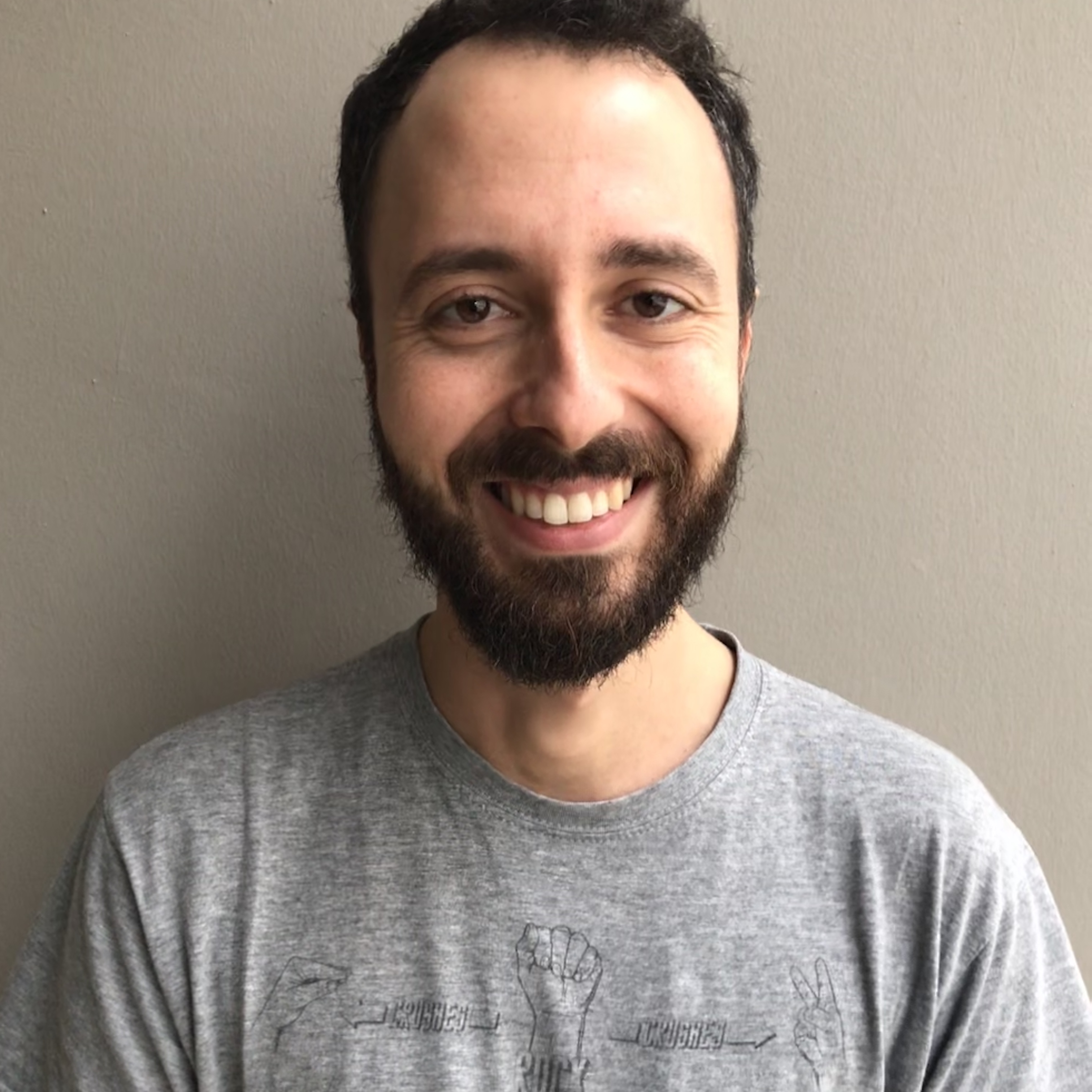}}]{Claudio Pomo}
is an assistant professor at Politecnico di Bari focused on responsible AI for personalization, reproducibility, and multi-objective evaluation. His work appears at SIGIR, RecSys, ECIR, UMAP, and in Information Science and IP\&M; he co-led tutorials (RecSys 2021; LoG 2023), organized the EvalRS workshop at KDD 2023, chaired the RecSys Challenge 2024–2025, and lectured at the RecSys Summer School 2024.
\end{IEEEbiography}

%\vspace{-4em}

\begin{IEEEbiography}[{\includegraphics[width=1in,height=1.25in,clip,keepaspectratio]{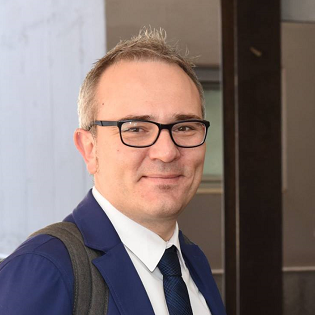}}]{Tommaso Di Noia}
a Professor of Computer Science at the Polytechnic University of Bari, Italy. His work spans AI and data management, evolving from knowledge representation and automated reasoning to automated negotiation and recommender systems with knowledge graphs and Linked Open Data. He now focuses on trustworthy AI—adversarial ML, explainability, fairness, and privacy in recommender systems—and served as program co-chair of RecSys 2023 and general co-chair of RecSys 2024.
% is a Professor of Computer Science at Polytechnic University of Bari, Italy. His research activities focus on AI and Data Management. They were initially devoted to knowledge representation and automated reasoning. Then, he studied how to apply knowledge representation techniques to automated negotiations. Following these ideas, he has devoted his interest to applying knowledge graphs and Linked Open Data to RSs with papers published in international journals, conferences, and book chapters. During the last years, he moved his research into the Trustworthy AI topic with a particular interest in adversarial ML, explainability, fairness, and privacy protection of RSs. He is serving as program co-chair at RecSys 2023 and general co-chair at RecSys 2024.
\end{IEEEbiography}

%\vspace{-4em}

\begin{IEEEbiography}[{\includegraphics[width=1in,height=1.25in,clip,keepaspectratio]{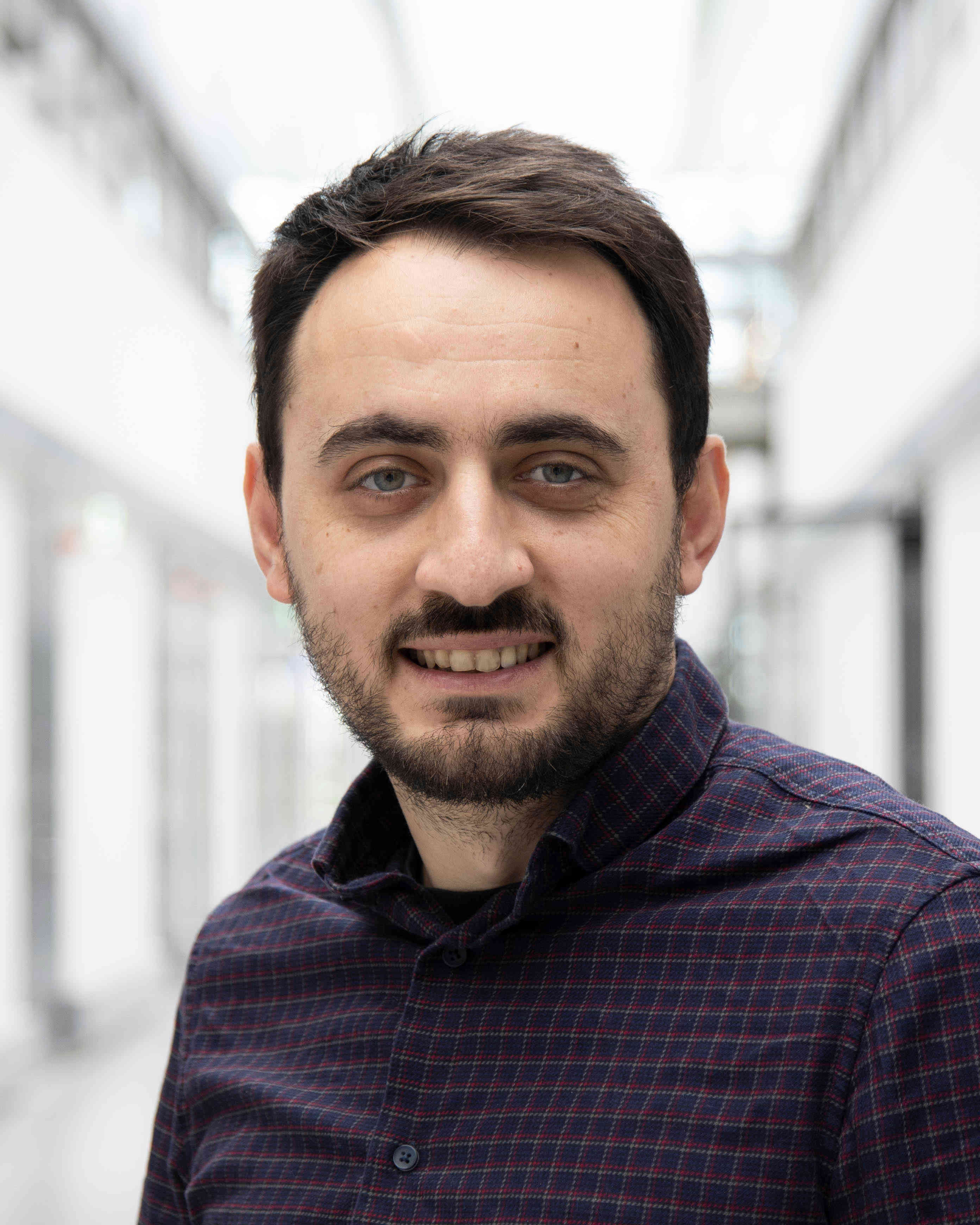}}]{Fragkiskos D. Malliaros}
is a Professor at Université Paris-Saclay, CentraleSupélec, Inria. Previously, he was a postdoctoral researcher at UC San Diego (2016-17) and École Polytechnique (2015-16). He is the recipient of the 2012 Google European Doctoral Fellowship in Graph Mining, the 2015 Thesis Prize by École Polytechnique, and best paper awards at TextGraphs-NAACL 2018 and AAAI ICWSM 2020 (honorable mention). His current research interests focus on graph machine learning and applications.
\end{IEEEbiography}

\end{document}